\documentclass[10pt,a4paper]{article}
\usepackage[utf8]{inputenc}
\usepackage[english]{babel}

\usepackage{amsmath}
\usepackage{amsfonts}
\usepackage{amssymb}

\usepackage[colorlinks,citecolor=blue,urlcolor=blue,linkcolor=blue]{hyperref}

\usepackage[left=2cm,right=2cm,top=2cm,bottom=2cm]{geometry}

\usepackage{authblk}

\usepackage{amsthm}
\newtheorem{definition}{Definition}
\newtheorem{corollary}{Corollary}
\newtheorem{theorem}{Theorem}

\usepackage{feynmp}
\newcommand{\pd}{\partial}
\newcommand{\tr}{\operatorname{tr}}

\usepackage{listings}

\title{FeynGrav : FeynCalc extension for gravity amplitudes}
\author[1,2]{\href{https://orcid.org/0000-0001-7099-0861}{B. Latosh} \thanks{\href{mailto:latosh@theor.jinr.r}{latosh@theor.jinr.ru}}}
\affil[1]{Bogoliubov Laboratory of Theoretical Physics, JINR, Dubna 141980, Russia}
\affil[2]{Dubna State University, Universitetskaya str. 19, Dubna 141982, Russia}
\date{}

\begin{document}

\maketitle

\begin{abstract}
  Package ``FeynGrav'' which provides a framework to operate with Feynman rules for gravity within ``FeynCalc'' is presented. We present a framework to deal with Feynman rules for general relativity and non-supersymmetric matter minimally coupled to gravity. Applicability of the package is tested with $2\to2$ on-shell tree level graviton scattering, polarization operators, and one-loop scalar-gravitational interaction structure.
\end{abstract}

\section{Introduction}

Feynman rules for gravity are notoriously hard to derive because of the large number of terms constituting expressions for each vertex. As it is noted in \cite{DeWitt:1967yk,DeWitt:1967ub,DeWitt:1967uc}, in one particular parameterization an expression for the three graviton vertex contains at least $171$ terms, while an expression for the four graviton vertex contains $2850$ terms. Despite this complexity there are quite a few publications where such rules are used \cite{DeWitt:1967yk,DeWitt:1967ub,DeWitt:1967uc,Grisaru:1975ei,Goroff:1985th,Donoghue:1994dn,Akhundov:1996jd,BjerrumBohr:2002kt,Holstein:2006bh,Jakobsen:2020diz,Prinz:2020nru}. We would especially like to highlight the recent publication \cite{Prinz:2020nru} where explicit analytic formulae for gravitational interactions for any valency were obtained.

Despite the fact that Feynman rules for gravity are presented in literature, still, they exist in a form barely suitable for manual calculations because of their length and complexity. Calculations presented in paper \cite{Sannan:1986tz} serve as a suitable illustration. The paper is devoted to $2 \to 2$ on-shell graviton scattering at the tree level. To calculate the corresponding matrix element the author operates with special symmetrized expressions. Moreover, all external gravitons are fixed on-shell, so the corresponding polarization operators provide an additional mean to simplify the expression. But even after such a simplification the resulted calculations are about a page long.

Difficulties that come alone with the large number of terms can be soften with contemporary computer algebra packages because expressions containing hundreds or thousands of terms can be (relatively) easily manipulated by a computer. A detailed review of contemporary software tools used in gravity and high energy physics lies beyond the scope of this paper and can be found in \cite{MacCallum:2018csx,HEPSoftwareFoundation:2020daq}. In the context of the present work we shall highlight two pieces of software. The first one is ``FeynCalc'' which provide a set of tools to operate within matrix elements within quantum field theory \cite{Shtabovenko:2016sxi,Shtabovenko:2020gxv,Mertig:1990an}. The second one is ``xAct'' which provide a framework to operate with various models of gravity \cite{xActBundle,Portugal:1998qi,MARTINGARCIA2008597,Brizuela:2008ra,Pitrou:2013hga,Nutma:2013zea}. We will use FeynCalc as a basis for a framework that generates Feynman rules for gravity. On the contrary, xAct is not used in this paper, but it provides a framework that may be implemented for future research. We will return to a discussion of both FeynCalc and xAct later.

This paper has two objectives. The first one is to present a (relatively) simple algorithm to obtain expressions for Feynman rules for gravity in a form suitable for implementation in a computer algebra system. The second one is to present ``FeynGrav'', an extension for FeynCalc package that implements the given algorithm. We obtain Feynman rules for general relativity and for non-supersymmetric minimally coupled scalars, massless vectors, and Dirac fermions. We believe that the discussed mathematical framework also provides a way to obtain Feynman rules in other models of gravity.

The paper is organized as follows. The most part of the paper is devoted to a discussion of the derivation of Feynman rules. In Section \ref{Section_tensor_framework_bosons} we define the tensor framework that is used throughout the paper. Namely, we define $I$ and $C$ tensors which provide a tool to make the tensors structure of bosonic Lagrangians explicit. In Section \ref{Section_GR_rules} we describe how rules for pure general relativity are obtained within that framework. In Section \ref{Section_Scalars} we derive Feynman rules for a massive scalar field minimally coupled to gravity. In Section \ref{Section_Vectors} we derive Feynman rules for a massless vector field minimally coupled to gravity. In Section \ref{Section_Fermions} we discuss the standard way to generalize the Dirac action for the curved spacetime case and derive the corresponding Feynman rules. In Section \ref{Section_Implementation} we discuss FeynGrav package which implements the presented calculations in FeynCalc. We show a few examples where FeynGrav serve as a useful tool. Finally, we conclude the discussion in Section \ref{Section_Conclusion} and make a few comment on applicability and future development of FeynGrav.

\section{Tensor framework}\label{Section_tensor_framework_bosons}

Feynman rules for gravity are obtained within a framework of perturbative effective gravity. This framework is based on the following assumptions. Firstly, the model under consideration is assumed to be applicable only below the Planck scale. Secondly, the gravity is associated with small metric perturbations $h_{\mu\nu}$ about the flat background spacetime $\eta_{\mu\nu}$. To some extend such a theory can be viewed as a gauge theory of massless rank-$2$ symmetric tensor $h_{\mu\nu}$.

The complete spacetime metric is composed from both background and perturbations:
\begin{align}\label{The_perturbative_expansion}
  g_{\mu\nu} = \eta_{\mu\nu} + \kappa \, h_{\mu\nu} \, .
\end{align}
Here $\kappa$ is related with the Newton constant $G_N$ as follows:
\begin{align}
  \kappa^2 =32 \pi\, G_N .
\end{align}
Expression \eqref{The_perturbative_expansion} is exact and does not contain terms of higher orders in $\kappa$. Other geometric quantities, for instance, $\sqrt{-g}$ and $g^{\mu\nu}$, are infinite series in $\kappa$. The tensor framework defined further is designed to make the tensor structure of such expansions explicit.

\begin{definition}
  {~}\\
  $I_{\mu_1\nu_1\cdots\mu_n\nu_n}$ tensor of the $n$-the order has $n$ pairs of indices. $I_{\mu_1\nu_1\mu_2\nu_2\cdots \mu_n\nu_n}$ is obtained from $\eta_{\nu_1\mu_2}\eta_{\mu_2\nu_3}\cdots\eta_{\nu_n\mu_1}$ by a consecutive symmetrization with respect to indices in each pair $(\mu_i,\nu_i)$.
\end{definition}

Following comments on the definition are due. Firstly, the given definition is symmetric with respect to permutations of indices within each index pair $(\mu_i,\nu_i)$, but it is not symmetric with respect to permutations of such pairs. This is a deliberate decision as for the most part of calculations it is not required to operate with expressions enjoying such a symmetry. Secondly, the symmetrization with respect to $\mu_i\leftrightarrow \nu_i$ is assumed to be performed with the factor $1/2$:
\begin{align}
  \eta_{\nu_1\mu_2}\eta_{\nu_2\mu_3}\cdots \eta_{\nu_n\mu_1} &\to \cfrac12 \left[ \eta_{\nu_1\mu_2}\eta_{\nu_2\mu_3}\cdots \eta_{\nu_n\mu_1}+[\mu_1\leftrightarrow \nu_1] \right]\\
  &\to\cfrac12\,\left(\, \cfrac12 \left( \eta_{\nu_1\mu_2}\eta_{\nu_2\mu_3}\cdots \eta_{\nu_n\mu_1}+[\mu_1\leftrightarrow \nu_1] \right)+[\mu_2\leftrightarrow\nu_2 ] \right)\nonumber \\
  &\hspace{5pt}\vdots\nonumber \\
  &\to I_{\mu_1\nu_1\mu_2\nu_2\cdots\mu_n\nu_n} .\nonumber
\end{align}
\noindent For the sake of illustration we present $I$ tensors of first four orders.
\begin{corollary}
  \begin{align}\label{I-tensors_definition}
    I^{\mu\nu} =& \eta^{\mu\nu} \, , \\
    I^{\mu\nu\alpha\beta} =& \cfrac12\, \left[ \eta^{\mu\alpha}\eta^{\nu\beta} + \eta^{\mu\beta}\eta^{\nu\alpha} \right]\, , \nonumber\\
    I^{\mu\nu\alpha\beta\rho\sigma} =& \cfrac18 \Bigg[ \eta^{\alpha\sigma} \eta^{\beta\nu}\eta^{\mu\rho} + \eta^{\alpha\nu}\eta^{\beta\sigma}\eta^{\mu\rho} + \eta^{\alpha\rho}\eta^{\beta\nu}\eta^{\mu\sigma} + \eta^{\alpha\nu}\eta^{\beta\rho}\eta^{\mu\sigma}\nonumber \\
      &\hspace{10pt}+\eta^{\alpha\sigma}\eta^{\beta\mu}\eta^{\nu\rho}+\eta^{\alpha\mu}\eta^{\beta\sigma}\eta^{\nu\rho}+\eta^{\alpha\rho}\eta^{\beta\mu}\eta^{\nu\sigma}+\eta^{\alpha\mu}\eta^{\beta\rho}\eta^{\nu\sigma} \Bigg]\nonumber,\\
    I^{\mu\nu\alpha\beta\rho\sigma\lambda\tau} =&\cfrac{1}{16}\Bigg[ \eta^{\alpha\sigma}\eta^{\beta\nu}\eta^{\lambda\rho}\eta^{\mu\tau} + \eta^{\alpha\nu}\eta^{\beta\sigma}\eta^{\lambda\rho}\eta^{\mu\tau} + \eta^{\alpha\rho}\eta^{\beta\nu}\eta^{\lambda\sigma}\eta^{\mu\tau} + \eta^{\alpha\nu}\eta^{\beta\rho}\eta^{\lambda\sigma}\eta^{\mu\tau} + \eta^{\alpha\sigma}\eta^{\beta\mu}\eta^{\lambda\rho}\eta^{\nu\tau} \nonumber \\
      &\hspace{12pt} +\eta^{\alpha\mu}\eta^{\beta\sigma}\eta^{\lambda\rho}\eta^{\nu\tau} + \eta^{\alpha\rho}\eta^{\beta\mu}\eta^{\lambda\sigma}\eta^{\nu\tau} + \eta^{\alpha\mu}\eta^{\beta\rho}\eta^{\lambda\sigma}\eta^{\nu\tau} + \eta^{\alpha\sigma}\eta^{\beta\nu}\eta^{\lambda\mu}\eta^{\rho\tau} + \eta^{\alpha\nu}\eta^{\beta\sigma}\eta^{\lambda\mu}\eta^{\rho\tau} \nonumber \\
      &\hspace{12pt} +\eta^{\alpha\sigma}\eta^{\beta\mu}\eta^{\lambda\nu}\eta^{\rho\tau}+\eta^{\alpha\mu}\eta^{\beta\sigma}\eta^{\lambda\nu}\eta^{\rho\tau}+\eta^{\alpha\rho}\eta^{\beta\nu}\eta^{\lambda\mu}\eta^{\sigma\tau}+\eta^{\alpha\nu}\eta^{\beta\rho}\eta^{\lambda\mu}\eta^{\sigma\tau}+\eta^{\alpha\rho}\eta^{\beta\mu}\eta^{\lambda\nu}\eta^{\sigma\tau} \nonumber\\
      &\hspace{12pt}+\eta^{\alpha\mu}\eta^{\beta\rho}\eta^{\lambda\nu}\eta^{\sigma\tau} \Bigg] \,.\nonumber
  \end{align}
\end{corollary}
\noindent Although it is hard to manually perform such a symmetrization it can be (relatively) easily implemented within a computer algebra system.

The given definition of $I$ tensors provides a way to separate the tensor structure of $(h^n)^{\mu\nu}$ and $\tr(h^n)$.
\begin{corollary}
  If $h_{\mu\nu}$ is an arbitrary symmetric tensor then
  \begin{align}
    (h^n)^{\mu\nu} \overset{\text{note}}{=}h^{\mu}{}_{\sigma_1} h^{\sigma_1}{}_{\sigma_2} \cdots h^{\sigma_{n-1} \nu} = I^{\mu\nu\alpha_1\beta_1\cdots\alpha_n \beta_n} h_{\alpha_1\beta_1} \cdots h_{\alpha_n\beta_n} \,.
  \end{align}
\end{corollary}

\begin{corollary}
  If $h_{\mu\nu}$ is an arbitrary symmetric tensor then
  \begin{align}
    \tr (h^n) \overset{\text{note}}{=} h^{\sigma_1}{}_{\sigma_2} h^{\sigma_2}{}_{\sigma_3} \cdots h^{\sigma_n}{}_{\sigma_1} = I^{\alpha_1\beta_1\cdots \alpha_n \beta_n} h_{\alpha_1\beta_1} h_{\alpha_2\beta_2} \cdots h_{\alpha_n\beta_n}\,.
  \end{align}
\end{corollary}
\noindent This property allows one to use $I$ tensors to separate the tensor structure of any expression that contains $h_{\mu\nu}$ and free from derivatives.

First and foremost, let us construct an expansion for $g^{\mu\nu}$. As it is noted above, $g_{\mu\nu}$ is not a series but a finite expression \eqref{The_perturbative_expansion}. The inverse metric $g^{\mu\nu}$, in turn, is an infinite series given by the following expression\footnote{In this paper we do not discuss the convergence of any such series. Firstly, the subject lies beyond the scope of this paper. Secondly, we believe it is covered well enough in the aforementioned paper \cite{Prinz:2020nru}.}
\begin{align}
  g^{\mu\nu} = \eta^{\mu\nu} - \kappa\, h^{\mu\nu} + \kappa^2 \,h^\mu{}_\sigma h^{\sigma\nu} + \cdots = \sum\limits_{n=0}^\infty \, (-\kappa)^n (h^n)^{\mu\nu} \,.
\end{align}
In terms of $I$ tensors this series takes the following form.
\begin{theorem}
  \begin{align}\label{inverse_g_expansion}
    g^{\mu\nu} = \sum\limits_{n=0}^\infty \, (-\kappa)^n \, I^{\mu\nu \,\rho_1\sigma_1\cdots\rho_n\sigma_n} \, h_{\rho_1\sigma_1}\cdots h_{\rho_n \sigma_n}
  \end{align}
\end{theorem}

The tensor structure of $\sqrt{-g}$ is recovered in a similar way. The factor is expanded in the following infinite series
\begin{align}
  \sqrt{-g} =& \exp\left[\cfrac12\,\tr\left\{\ln \left(\delta_\mu^\nu + \kappa\, h_\mu{}^\nu\right)\right\} \right]=\exp\left[-\cfrac12\,\sum\limits_{n=1}^\infty\cfrac{(-\kappa)^n}{n} \tr \left(h^n\right) \right] \,.
\end{align}
Let us examine the structure of this expansion. Factor $\sqrt{-g}$ is as an exponent of an infinite series, therefore it is an infinite series each term of which is a power of another infinite series:
\begin{align}
  \sqrt{-g} =&\exp\left[-\cfrac12\,\sum\limits_{n=1}^\infty\cfrac{(-\kappa)^n}{n} \tr \left(h^n\right) \right] \\
  =& 1 + \cfrac{1}{1!} \left(-\cfrac12\,\sum\limits_{p=1}^\infty\cfrac{(-\kappa)^p}{p} \tr \left(h^p\right)\right)^1 +\cdots +\cfrac{1}{n!} \left(-\cfrac12\,\sum\limits_{p=1}^\infty\cfrac{(-\kappa)^p}{p} \tr \left(h^p\right)\right)^n +\cdots\nonumber \,.
\end{align}
The first term of the series is the only background term, i.e. it is proportional to $\kappa^0$. Each consequent term contains an infinite number of powers of $\kappa$. However the second term starts with $\kappa^1$, the third term starts with $\kappa^2$, and so on. Because of this contributions proportional to $\kappa^n$ are contained only in first $n+1$ terms. Therefore, at order $\mathcal{O}(\kappa^n)$ the series contains a finite number of terms and the corresponding contribution is given by the following theorem.
\begin{theorem}
  \begin{align}\label{volume_measure_expansion_a}
    \left. \sqrt{-g}\right|_{\mathcal{O}(\kappa^n)} =& (-\kappa)^n \,\sum\limits_{m=1}^n \cfrac{1}{m!} \left(-\cfrac12\right)^m\left[\sum\limits_{k_1+\cdots +k_m=n}\,\cfrac{1}{k_1 \times\cdots\times k_m} \tr(h^{k_1}) \cdots \tr(h^{k_m})\right]\\
    =& \kappa^n \, h_{\mu_1\nu_1} \cdots h_{\mu_n\nu_n} \left[\sum\limits_{m=1}^n \cfrac{(-1)^{n+m}}{m! 2^m}\sum\limits_{k_1+\cdots+k_m=n}\,\cfrac{1}{k_1\times\cdots\times k_m} \, \left(I_{(k_1)} \cdots I_{(k_m)}\right)^{\mu_1\nu_1\cdots\mu_n\nu_n}\right].\nonumber
  \end{align}
  Here $\left(I_{(k_1)} \cdots I_{(k_m)}\right)^{\mu_1\nu_1\cdots\mu_n\nu_n}$ denotes a multiplication of $m$ exemplars of $I$ tensors\footnote{For example, $\left(I_{(1)}I_{(2)}I_{(3)}\right)^{\mu_1\nu_1\cdots\mu_6\nu_6} = I^{\mu_1\nu_1} I^{\mu_2\nu_2\mu_3\nu_3} I^{\mu_4\nu_4\mu_5\nu_5\mu_6\nu_6}$}. The first tensor has $k_1$ pairs of indices, the second one has $k_2$, and so on. And the overall index order is $\mu_1,\nu_1,\mu_2,\nu_2,\cdots,\mu_n,\nu_n$.
\end{theorem}
\noindent Formula \eqref{volume_measure_expansion_a} states that in order to obtain $\sqrt{-g}\big|_{\mathcal{O}(\kappa^n)}$ one takes a singe term from the second term of the exponent expansion, two terms from the third term of the exponent expansion and so on.

It must be noted that in an analogy with $I$ tensors the following expression
\begin{align}
  \sum\limits_{m=1}^n \cfrac{(-1)^{n+m}}{m! 2^m}\sum\limits_{k_1+\cdots+k_m=n}\,\cfrac{1}{k_1\cdots k_m} \, \left(I_{(k_1)} \cdots I_{(k_m)}\right)^{\mu_1\nu_1\cdots\mu_n\nu_n}
\end{align}
is symmetric with respect to indices within each index pair $(\mu_i,\nu_i)$, but not symmetric with respect to pairs transmutations $(\mu_i,\nu_i)\leftrightarrow(\mu_j,\nu_j)$. For the sake of convenience we perform that symmetrization in the following definition.

\begin{definition}
  {~}\\
  $C$ tensor of the $n$-the order has $2n$ indices. C tensor is defined as follows:
  \begin{align}\label{C-tensors_definition}
    \mathcal{C}^{\mu_1\nu_1\cdots\mu_n\nu_n}=\operatorname{Symm} \left[\sum\limits_{m=1}^n \cfrac{(-1)^{n+m}}{m! 2^m}\sum\limits_{k_1+\cdots+k_m=n}\,\cfrac{1}{k_1\cdots k_m} \, \left(I_{(k_1)} \cdots I_{(k_m)}\right)^{\mu_1\nu_1\cdots\mu_n\nu_n}\, \right].
  \end{align}
  Here $\operatorname{Symm}$ notes a symmetrization with respect to all possible permutations of pairs $(\mu_i,\nu_i)\leftrightarrow(\mu_j,\nu_j)$. By the symmetrization we mean that a summation over all permutation of index pairs must be performed and the results is {\bf divided} by the overall number of permutations.
\end{definition}
\noindent Let us present a few $C$ tensors for the sake of illustration.
\begin{corollary}
  \begin{align}
    \mathcal{C}^{\mu\nu} =& \cfrac12 \, \eta^{\mu\nu} ,\\
    \mathcal{C}^{\mu\nu\alpha\beta} =& \cfrac18 \Bigg[ -\eta^{\alpha\nu}\eta^{\beta\mu}-\eta^{\alpha\mu}\eta^{\beta\nu}+\eta^{\alpha\beta}\eta^{\mu\nu} \Bigg] ,\nonumber \\
    \mathcal{C}^{\mu\nu\alpha\beta\rho\sigma} =&\cfrac{1}{48} \Bigg[ -\eta^{\alpha\sigma}\eta^{\beta\rho}\eta^{\mu\nu} - \eta^{\alpha\rho}\eta^{\beta\sigma}\eta^{\mu\nu} + \eta^{\alpha\beta}\eta^{\mu\nu}\eta^{\rho\sigma} + \eta^{\alpha\sigma}\eta^{\beta\nu}\eta^{\mu\rho} + \eta^{\alpha\nu}\eta^{\beta\sigma}\eta^{\mu\rho}\nonumber\\
      &\hspace{20pt} + \eta^{\alpha\rho}\eta^{\beta\nu}\eta^{\mu\sigma} +\eta^{\alpha\nu}\eta^{\beta\rho}\eta^{\mu\sigma} + \eta^{\alpha\sigma}\eta^{\beta\mu}\eta^{\nu\rho} + \eta^{\alpha\mu}\eta^{\beta\sigma}\eta^{\nu\rho} - \eta^{\alpha\beta}\eta^{\mu\sigma}\eta^{\nu\rho} \nonumber \\
      &\hspace{20pt} + \eta^{\alpha\rho}\eta^{\beta\mu}\eta^{\nu\sigma} +\eta^{\alpha\mu}\eta^{\beta\rho}\eta^{\nu\sigma}-\eta^{\alpha\beta}\eta^{\mu\rho}\eta^{\nu\sigma}-\eta^{\alpha\nu}\eta^{\beta\mu}\eta^{\rho\sigma}-\eta^{\alpha\mu}\eta^{\beta\nu}\eta^{\rho\sigma} \Bigg] . \nonumber 
  \end{align}
\end{corollary}

\begin{theorem}
  \begin{align}\label{formula_m1}
    \sqrt{-g} = \sum\limits_{n=0}^\infty \, \kappa^n \,\mathcal{C}^{\mu_1\nu_1\cdots\mu_n\nu_n}\, h_{\mu_1\nu_1} \cdots h_{\mu_n\nu_n} .
  \end{align}
\end{theorem}

Given definitions of $I$ and $C$ tensors provide a direct way to separate the tensor structure of the following expressions.

\begin{theorem}
  \begin{align}
    \sqrt{-g} \, g^{\mu\nu} &=\sum\limits_{n=0}^\infty \, \kappa^n\, \mathcal{C}_I^{\mu\nu|\rho_1\sigma_1\cdots\rho_n\sigma_n} \, h_{\rho_1\sigma_1} \cdots h_{\rho_n\sigma_n}\,,\\
    \sqrt{-g} \, g^{\mu\nu} g^{\alpha\beta} &= \sum\limits_{n=0}^\infty \, \kappa^n\, \mathcal{C}_{II}^{\mu\nu\alpha\beta|\rho_1\sigma_1\cdots\rho_n\sigma_n} \, h_{\rho_1\sigma_1} \cdots h_{\rho_n\sigma_n}\,,\nonumber \\
    \sqrt{-g} \, g^{\mu\nu} g^{\alpha\beta} g^{\lambda\tau} &= \sum\limits_{n=0}^\infty \, \kappa^n\, \mathcal{C}_{III}^{\mu\nu\alpha\beta\lambda\tau|\rho_1\sigma_1\cdots\rho_n\sigma_n} \, h_{\rho_1\sigma_1} \cdots h_{\rho_n\sigma_n}\nonumber,
  \end{align}
  with
  \begin{align}
    \mathcal{C}_I^{\mu\nu|\rho_1\sigma_1\cdots\rho_n\sigma_n} =&\operatorname{Symm}\left[\mathop{ \sum\limits_{n_0,n_1=0}^n}_{n_0+n_1=n} (-1)^{n-n_0} \left( I_{(n_1+1)}^{\mu\nu \cdots}  \mathcal{C}_{(n_0)}^{\cdots} \right)^{\rho_1\sigma_1\cdots \rho_n\sigma_n}\right]\,,\\
    \mathcal{C}_{II}^{\mu\nu\alpha\beta|\rho_1\sigma_1\cdots\rho_n\sigma_n}=&\operatorname{Symm}\left[ \mathop{ \sum\limits_{n_0,n_1,n_2=0}^n}_{n_0+n_1+n_2=n} (-1)^{n-n_0} \left( I_{(n_1+1)}^{\mu\nu\cdots} I_{(n_2+1)}^{\alpha\beta\cdots} \, \mathcal{C}_{(n_0)}^{\cdots} \right)^{\rho_1\sigma_1\cdots\rho_n\sigma_n}\right]\,,\nonumber \\
    \mathcal{C}_{III}^{\mu\nu\alpha\beta\lambda\tau|\rho_1\sigma_1\cdots\rho_n\sigma_n} =&\operatorname{Symm}\left[\mathop{ \sum\limits_{n_0,n_1,n_2,n_3=0}^n }_{n_0+n_1+n_2+n_3=n}\hspace{-20pt} (-1)^{n-n_0} \,\left( I_{(n_1+1)}^{\mu\nu\cdots} I_{(n_2+1)}^{\alpha\beta\cdots} I_{(n_3+1)}^{\lambda\tau\cdots} \, \mathcal{C}_{(n_0)}^{\cdots} \right)^{\rho_1\sigma_1\cdots\rho_n\sigma_n} \right]\,.\nonumber
  \end{align}
  Here $\operatorname{Symm}$ notes a symmetrization with respect to all possible permutations of $(\rho_i,\sigma_i)$ pairs. Expressions with I and C tensors taken in brackets note a multiplication of the corresponding tensors with the external indices consecutively replacing the dots. For instance:
  \begin{align}
    \left( I_{(1+0)}^{\mu\nu\cdots} I_{(1+1)}^{\alpha\beta\cdots} I_{(1+2)}^{\lambda\tau\cdots} \, \mathcal{C}_{3}^{\cdots} \right)^{\rho_1\sigma_1\cdots\rho_6\sigma_6}\!\! = I^{\mu\nu} I^{\alpha\beta\rho_1\sigma_1} I^{\lambda\tau\rho_2\sigma_2\rho_3\sigma_3}\mathcal{C}^{\rho_4\sigma_4\rho_5\sigma_5\rho_6\sigma_6}\,.
  \end{align}
\end{theorem}

These factors appear in bosonic Lagrangians that are discussed below. They are sufficient to obtain perturbative expansions for general relativity, scalar, and vector fields. Description of fermions, however, is much more sophisticated and presented separately in Section \ref{Section_Fermions}.

\section{General relativity}\label{Section_GR_rules}

General relativity is given by the well known action
\begin{align}
  \mathcal{A}_{s=2,m=0} =& \int d^4 x \sqrt{-g} \left[-\cfrac{2}{\kappa^2} \, R\right].\label{Hilbert_action}
\end{align}
For the sake of distinguishment we call \eqref{Hilbert_action} the Hilbert parameterization. This parameterization is not suitable for perturbative treatment. It shows that gravity is, indeed, a geometry of a curved spacetime, but its perturbative structure remains implicit. To make the perturbative structure explicit a full derivative shall be separated from \eqref{Hilbert_action}.

\begin{theorem}
  \begin{align}
    \sqrt{-g} R =&~\cfrac14\, \sqrt{-g} \, g^{\mu\nu}g^{\alpha\beta}g^{\rho\sigma}\left[\pd_\mu g_{\alpha\beta}\,\pd_\nu g_{\rho\sigma} -\pd_\mu g_{\alpha\rho}\, \pd_\nu g_{\beta\sigma}+2\,\pd_\mu g_{\beta\rho}\, \pd_\alpha g_{\nu\sigma} - 2\, \pd_\mu g_{\nu\alpha}\, \pd_\beta g_{\rho\sigma}\right]\\
    &~+\text{full derivative} \, .\nonumber
  \end{align}
\end{theorem}
\noindent To prove this theorem one should, firstly, express the scalar curvature via Christoffel symbols and, secondly, perform an integration by parts:
\begin{align}
  \sqrt{-g} \, R = \sqrt{-g} \,g^{\mu\nu} \left[\Gamma^\sigma_{\mu\rho}\Gamma^\rho_{\nu\sigma} - \Gamma^\sigma_{\mu\nu} \Gamma^\rho_{\sigma\rho}\right] + \pd_\mu \left( \sqrt{-g}\, g^{\alpha\beta}\,\Gamma^\mu_{\alpha\beta}  -\sqrt{-g}\, g^{\mu\nu}\,\Gamma^\sigma_{\nu\sigma}\right).
\end{align}
Lastly, Christoffel symbols should be expressed via the metric:
\begin{align}
  \Gamma^\alpha_{\mu\nu} = \cfrac12\,g^{\alpha\beta} \left[\pd_\mu g_{\beta\nu} + \pd_\nu g_{\beta\mu} - \pd_\beta g_{\mu\nu}\right].
\end{align}
We shall note that this theorem is well known \cite{Dubrovin_Fomenko_Novikov,Novikov_Taimanov}, so we do not discuss its proof in details. From here on we omit all surface terms.

This theorem provides a way to write action \eqref{Hilbert_action} in the following form which we call the Einstein parameterization:
\begin{align}\label{Einstein_action}
  \mathcal{A}_{s=2,m=0}\! =\!\!\!\int d^4 x\sqrt{-g} g^{\mu\nu}g^{\alpha\beta}g^{\rho\sigma}\left[-\cfrac{1}{2\kappa^2}\left(\pd_\mu g_{\alpha\beta}\,\pd_\nu g_{\rho\sigma} -\pd_\mu g_{\alpha\rho}\, \pd_\nu g_{\beta\sigma}+2\,\pd_\mu g_{\beta\rho}\, \pd_\alpha g_{\nu\sigma} - 2\, \pd_\mu g_{\nu\alpha}\, \pd_\beta g_{\rho\sigma}\right)\right].
\end{align}
For the perturbative expansion \eqref{The_perturbative_expansion} this formula is reduced further because the background metric has vanishing derivatives:
\begin{align}\label{Einstein_perturbative_action}
  \mathcal{A}_{s=2,m=0} =\int d^4 x\sqrt{-g}\, g^{\mu\nu}g^{\alpha\beta}g^{\rho\sigma}\left[-\cfrac{1}{2}\left(\pd_\mu h_{\alpha\beta}\pd_\nu h_{\rho\sigma} -\pd_\mu h_{\alpha\rho}\pd_\nu h_{\beta\sigma}+2\,\pd_\mu h_{\beta\rho} \pd_\alpha h_{\nu\sigma} - 2\, \pd_\mu h_{\nu\alpha} \pd_\beta h_{\rho\sigma}\right)\right].
\end{align}

Expression \eqref{Einstein_perturbative_action} shall be analyzed. The part of \eqref{Einstein_perturbative_action} placed in square brackets contains a finite number of term. The factor $\sqrt{-g}\, g^{\mu\nu} g^{\alpha\beta}g^{\rho\sigma}$ is free from derivatives and contains an infinite number of term, but its tensor structure is made explicit via $\mathcal{C}_{III}$ tensor. The tensor structure of terms placed in square brackets is factorized by the following theorem.

\begin{theorem}
  \begin{align}
    & \pd_\mu h_{\alpha\beta}\pd_\nu h_{\rho\sigma} -\pd_\mu h_{\alpha\rho}\pd_\nu h_{\beta\sigma}+2\,\pd_\mu h_{\beta\rho}\pd_\alpha h_{\nu\sigma} - 2\, \pd_\mu h_{\nu\alpha} \pd_\beta h_{\rho\sigma} \\
    &\to (p_1)^{\lambda_1} (p_2)^{\lambda_2} \, h{}^{\rho_1\sigma_1}_{(p_1)} \, h{}^{\rho_2\sigma_2}_{(p_2)}\, \mathcal{T}^{(h)}_{\mu\nu\alpha\beta\rho\sigma|\lambda_1\lambda_2\rho_1\sigma_1\rho_2\sigma_2} \,.\nonumber
  \end{align}
  with
  \begin{align}
    \mathcal{T}^{(h)}_{\mu\nu\alpha\beta\rho\sigma|\lambda_1\lambda_2\rho_1\sigma_1\rho_2\sigma_2}=& - \eta_{\mu\lambda_1}\eta_{\nu\lambda_2} I_{\alpha\beta\rho_1\sigma_1} I_{\rho\sigma\rho_2\sigma_2}+\eta_{\mu\lambda_1}\eta_{\nu\lambda_2} I_{\alpha\rho\rho_1\sigma_1} I_{\beta\sigma\rho_2\sigma_2}\nonumber\\
    & -2 \, \eta_{\mu\lambda_1}\eta_{\alpha\lambda_2} I_{\beta\rho\rho_1\sigma_1} I_{\nu\sigma\rho_2\sigma_2} +2 \, \eta_{\mu\lambda_1}\eta_{\beta\lambda_2} I_{\nu\alpha\rho_1\sigma_1} I_{\rho\sigma\rho_2\sigma_2} \,.
  \end{align}
\end{theorem}
\noindent Consequently, the full action \eqref{Einstein_perturbative_action} takes the following form.
\begin{theorem}
  \begin{align}\label{Einstein_perturbative_interaction_structure}
    \mathcal{A}_{s=2,m=0}=&\int\sum\limits_{n=0}^\infty \Bigg[ \cfrac{d^4 p_1}{(2\pi)^4}\cfrac{d^4 p_2}{(2\pi)^4}\prod\limits_{i=1}^n\cfrac{d^4 k_i}{(2\pi)^4} \, (2\pi)^4 \delta\left(p_1+p_2+\sum\limits_{i=1}^n k_i\right) \\
      &\times \kappa^n  \mathcal{C}_{III}^{\mu_1\nu_1\mu_2\nu_2\mu_3\nu_3|\rho_1\cdots\sigma_n}\,h{}_{\rho_1\sigma_1}^{(k_1)}\cdots h{}_{\rho_n\sigma_n}^{(k_n)}\,~\mathcal{T}^{(h)}_{\mu_1\nu_1\mu_2\nu_2\mu_3\nu_3|\lambda_1\lambda_2\alpha_1\beta_1\alpha_2\beta_2} \, (p_1)^{\lambda_1} (p_2)^{\lambda_2} \, h{}^{\alpha_1\beta_1}_{(p_1)} h^{\alpha_2\beta_2}_{(p_2)} \Bigg]. \nonumber
  \end{align}
\end{theorem}

This theorem does not yet provide expressions for the corresponding vertices as a certain symmetrization must be performed. Graviton vertices are quadratic in momenta due to the presence of derivatives. The corresponding expressions for interaction vertices are also quadratic in momenta, but they must be symmetric with respect to gravitons. Therefore the following symmetrization must be performed.

\begin{theorem}
  Expression $(\mathfrak{V}^{(n)})^{\rho_1\sigma_1\cdots\rho_n\sigma_n}$ that describes interaction of $n\geq 3$ gravitons reads
  \begin{align}
    (\mathfrak{V}^{(n)})^{\rho_1\sigma_1\cdots\rho_n\sigma_n}(p_1,\cdots,p_n) = i\,\kappa^n \left[\mathcal{C}_{III}^{\mu_1\nu_1\mu_2\nu_2\mu_3\nu_3|\rho_3\cdots\sigma_n}~\mathcal{T}^{(h)}_{\mu_1\nu_1\mu_2\nu_2\mu_3\nu_3|\lambda_1\lambda_2}{}^{\rho_1\sigma_1\rho_2\sigma_2} \, (p_1)^{\lambda_1} (p_2)^{\lambda_2} + \cdots\right].
  \end{align}
  Here $\cdots$ notes a summation over all possible permutations of $(p_i,\rho_i,\sigma_i)$. Therefore, there are $n!$ term in this sum; $\rho_i,\sigma_i$ are Lorentz indices of $i$-th graviton; $p_i$ is the corresponding graviton momentum.
\end{theorem}
\noindent This theorem provides a way to calculate rules for graviton interaction vertices at any order.

\section{Scalar field}\label{Section_Scalars}

A scalar field with a non-vanishing mass $m$ minimally coupled to gravity is described by the following action:
\begin{align}\label{scalar_action}
  \mathcal{A}_{s=0,m\not=0}=& \int d^4 x\sqrt{-g} \, \left[\cfrac12\,g^{\mu\nu} \nabla_\mu\phi \nabla_\nu\phi - \cfrac{m^2}{2} \, \phi^2\right] = \int d^4 x\left [ \sqrt{-g}\,g^{\mu\nu} \, \cfrac12 \, \pd_\mu\phi\, \pd_\nu\phi - \sqrt{-g}\, \cfrac{m^2}{2} \, \phi^2\right] .
\end{align}
It contain factors $\sqrt{-g}$ and $\sqrt{-g}\,g^{\mu\nu}$, but perturbative expansions for them are already obtained. Another part of the action also contains derivatives. The structure of this part in the momentum representation is given by the following theorem.
\begin{theorem}
  \begin{align}
    \cfrac12\,\pd_\mu\phi\, \pd_\nu\phi \to& \mathcal{T}{}^{(\phi)}_{\mu\nu}(p_1,p_2)\,\phi(p_1)\,\phi(p_2)
  \end{align}
  with
  \begin{align}
    \mathcal{T}{}^{(\phi)}_{\mu\nu}(p_1,p_2) =&-\cfrac12\,I_{\mu\nu\alpha\beta}\, (p_1)^\alpha (p_2)^\beta\,.
  \end{align}
\end{theorem}
\noindent Consequently, action \eqref{scalar_action} takes the following form.
\begin{theorem}
  \begin{align}
    \mathcal{A}_{s=0,m\not=0}=&\sum\limits_{n=0}^\infty\int \cfrac{d^4 p_1}{(2\pi)^4}\cfrac{d^4 p_2}{(2\pi)^4}\prod\limits_{i=1}^n\cfrac{d^4 k_i}{(2\pi)^4}\, (2\pi)^4 \delta\left(p_1+p_2+\sum\limits_{i=1}^n k_i\right) \\
    &\times \kappa^n\, \Bigg[  \mathcal{T}^{(\phi)}_{\mu\nu}(p_1,p_2)\, \mathcal{C}_I^{\mu\nu|\rho_1\cdots\sigma_n} - \cfrac{m^2}{2} \, \mathcal{C}^{\rho_1\cdots\sigma_n} \Bigg]\,h{}_{\rho_1\sigma_1}(k_1)\cdots h{}_{\rho_n\sigma_n}(k_n)\,\phi(p_1)\,\phi(p_2).\nonumber 
  \end{align}
\end{theorem}

The obtained expression by the construction is symmetric with respect both to gravitons and scalar, therefore no additional symmetrization shall be performed. At the same time an additional factor ``2'' shall be added. It appears because there are two equivalent ways to couple the given vertex to scalar lines. The corresponding Feynman rules are given by the following theorem.
\begin{theorem}
  Expressions $\mathfrak{V}^{(n)}_{s=0,m\not=0}$ that corresponds to a vertex describing interaction between two scalar particles and $n$ gravitons in given by the following expression:
  \begin{align}
    (\mathfrak{V}_{s=0,m\not =0}^{(n)}){}^{\rho_1\sigma_1\cdots\rho_n\sigma_n} (p_1,p_2)=&i\, \kappa^n\,2\, \left( \mathcal{T}^{(\phi)}_{\mu\nu}(p_1,p_2)\, \mathcal{C}_I^{\mu\nu|\rho_1\cdots\sigma_n} - \cfrac{m^2}{2} \,\mathcal{C}^{\rho_1\cdots\sigma_n} \right).
  \end{align}
\end{theorem}
\noindent The massless case is recovered in $m\to 0$ limit.

\section{Massless vector field}\label{Section_Vectors}

A single massless vector field is described by the following action:
\begin{align}\label{vector_action}
  \mathcal{A}_{s=1,m=0}=& \int d^4 x \sqrt{-g}\,g^{\mu\nu} g^{\alpha\beta} \, \left[-\cfrac14 \, F_{\mu\alpha} \, F_{\nu\beta}\right] .
\end{align}
The structure of $\sqrt{-g}\,g^{\mu\nu} g^{\alpha\beta}$ is obtained above. The structure of terms containing derivatives is given by the following theorem.
\begin{theorem}
  \begin{align}
    -\cfrac14\, F_{\mu\alpha} F_{\nu\beta} \to& \mathcal{T}{}^{(A)}_{\mu\nu\alpha\beta|\rho\sigma}(p_1,p_2)\,A^\rho(p_1)\,A^\sigma(p_2)
  \end{align}
  with
  \begin{align}
    \mathcal{T}^{(A)}_{\mu\nu\alpha\beta|\rho\sigma}(p_1,p_2)=\cfrac14\,(p_1)^\lambda (p_2)^\tau \left[I_{\mu\nu\rho\sigma}I_{\alpha\beta\lambda\tau}+I_{\mu\nu\lambda\tau}I_{\alpha\beta\rho\sigma}-I_{\mu\nu\rho\tau}I_{\alpha\beta\sigma\lambda}-I_{\mu\nu\lambda\sigma}I_{\alpha\beta\rho\tau}\right] .    
  \end{align}
\end{theorem}
\noindent Therefore the structure of \eqref{vector_action} is given by the next theorem.
\begin{theorem}
  \begin{align}
    \mathcal{A}_{s=1,m=0}=&\sum\limits_{n=0}^\infty \int\Bigg[ \cfrac{d^4 p_1}{(2\pi)^4}\cfrac{d^4 p_2}{(2\pi)^4}\prod\limits_{i=1}^n\cfrac{d^4 k_i}{(2\pi)^4} \, (2\pi)^4 \delta\left(p_1+p_2+\sum\limits_{i=1}^n k_i\right) \\
      &\times \kappa^n\, \mathcal{T}^{(A)}_{\mu\nu\alpha\beta|\lambda\tau}(p_1,p_2) \, \mathcal{C}_{II}^{\mu\nu\alpha\beta|\rho_1\cdots\sigma_n}\,h{}_{\rho_1\sigma_1}^{(k_1)}\cdots h{}_{\rho_n\sigma_n}^{(k_n)}\,A^\lambda(p_1)\,A^\tau(p_2) \Bigg] .\nonumber
  \end{align}
\end{theorem}

In full analogy with the previous case the obtained expression is symmetric both with respect to gravitons and vectors. An additional factor ``2'' shall also be recovered as there are two equivalent ways to couple the vertex to vector lines. The expression shall be used to obtain the Feynman rules without any additional symmetrization.

\begin{theorem}
  Expressions $\mathfrak{V}^{(n)}_{s=1,m=0}$ corresponding to a vertex describing interaction of two massless vector particles and $n$ gravitons in given by the following formula.
  \begin{align}
    (\mathfrak{V}_{s=1,m=0}^{(n)}){}_{\lambda_1\lambda_2|}{}^{\rho_1\sigma_1\cdots\rho_n\sigma_n}(p_1,p_2)=i\, \kappa^n\, 2\, \mathcal{T}^{(A)}_{\mu\nu\alpha\beta|\lambda\tau}(p_1,p_2) \mathcal{C}_{II}^{\mu\nu\alpha\beta|\rho_1\cdots\sigma_n} .
  \end{align}
\end{theorem}

A comment on the gauge fixing is due. As it was noted in Section \ref{Section_tensor_framework_bosons}, to a certain degree it is possible to consider the perturbative approach to gravity as a gauge theory of a rank $2$ symmetric tensor in a flat spacetime. Therefore the gauge of any matter field must be fixed after the perturbative expansions is made. The gauge of the gravitational field is fixed in accordance with that prescription, so shall be the gauge of a vector field. 

\section{Fermionic field}\label{Section_Fermions}

The standard way to describe fermions in a curved spacetime is based on vierbeins and $\gamma$ matrices \cite{Shapiro:2016pfm,stepanyants2009}. Firstly, $\gamma$ matrices are introduced to construct a representation of the Lorentz algebra. Dirac matrices are subjected to the following relations
\begin{align}
  \{ \gamma_m , \gamma_n \} \overset{\text{note}}{=} \gamma_m \gamma_n + \gamma_n \gamma_m &= 2\, \eta_{mn} \, , & \gamma^0 \gamma^m \gamma^0 & = (\gamma^m)^+ \,,
\end{align}
and form the following representation of the Lorentz algebra
\begin{align}\label{Lorentz_algebra_spinor_representation}
  S_{mn} & \overset{\text{def}}{=}  \cfrac{i}{4} \,[\gamma_m , \gamma_n] \, , &  [S_{mn}, S_{ab}] &= -i \left( \eta_{ma} S_{nb} - \eta_{mb} S_{na} +\eta_{nb} S_{ma} -\eta_{na} S_{mb} \right) .
\end{align}
The standard vector representation of the Lorentz algebra is constituted by matrices $J_{mn}$:
\begin{align}
  \left( J_{mn} \right)_{ab} = & \,i\, (\eta_{ma} \eta_{nb} - \eta_{mb} \eta_{na} ) \,, & [J_{mn}, J_{ab}] &= -i \left( \eta_{ma} J_{nb} - \eta_{mb} J_{na} +\eta_{nb} J_{ma} -\eta_{na} J_{mb} \right) .
\end{align}
These representations are connected by the following relation:
\begin{align}
  [\gamma_a , S_{mn} ] = \left( J_{mn}\right)_{ab} \,\gamma^b \,.
\end{align}
It allows one to define Dirac spinors $\psi$ and $\overline{\psi}=\psi^+ \gamma^0$ subjected to the following Lorentz group action
\begin{align}\label{spinor_transformations_flat_spacetime}
  \delta\psi &= \cfrac{i}{2} \, \omega^{ab}\, S_{ab}\, \psi  \, ,\\
  \delta\overline\psi &= -\cfrac{i}{2} \, \omega^{ab}\, \overline\psi S_{ab} \,,\nonumber
\end{align}
and to construct the well-known Lorentz invariant action (in a flat spacetime)
\begin{align}
  \mathcal{A}_{s=\frac12,m\not =0} = \int d^4 x \left[\cfrac{i}{2} \left(\, \overline\psi\, \gamma^m \pd_m \psi - \pd_m \overline\psi\, \gamma^m \psi \right) - m \,\overline\psi \, \psi\right].
\end{align}

Secondly, vierbeins $\mathfrak{e}_m{}^\mu$ are used to generalize this action for the curved spacetime case. In a curved spacetime one relates an arbitrary frame with a local inertial frame via vierbein $\mathfrak{e}_m{}^\mu$. Its Latin index is subjected to the Lorentz transformation while the Greek index is subjected to the general coordinate transformations. Vierbeins satisfy the following normalization condition:
\begin{align}
  \mathfrak{e}_m{}^\mu\, \mathfrak{e}_n{}^\nu\, g_{\mu\nu} =& \eta_{mn} , &\mathfrak{e}_m{}^\mu\, \mathfrak{e}_n{}^\nu\, \eta^{mn} =& g^{\mu\nu} .
\end{align}
They define anti-symmetric spin-connection $(\Gamma_\mu)_{ab} = - (\Gamma_\mu)_{ba}$ which is related with the Christoffel symbols:
\begin{align}\label{spin-connection}
  \left(\Gamma_\mu \right)_{ab} = \mathfrak{e}_{a\alpha} \mathfrak{e}_b{}^\beta \Gamma^\alpha_{\mu\beta} + \mathfrak{e}_{a\sigma}\pd_\mu\mathfrak{e}_b{}^\sigma .
\end{align}
Because vierbein $\mathfrak{e}_m{}^\mu$ relates Lorentz transformations and general coordinate transformations it shall be used to manipulate indices. This allows one to construct a generalization of $\gamma$ matrices subjected to the group of general coordinate transformations:
\begin{align}
  \gamma^\mu = \mathfrak{e}_m{}^\mu \,\gamma^m\,.
\end{align}
Spinor transformations \eqref{spinor_transformations_flat_spacetime} are generalized as follows
\begin{align}
  \delta\psi &= \cfrac{i}{2} \, \omega^{\mu\nu}\, S_{\mu\nu}\, \psi  = \cfrac{i}{2} \, \omega^{ab}(x)\, S_{ab}\, \psi \, ,\\
  \delta\overline\psi &= -\cfrac{i}{2} \, \omega^{\mu\nu}\, \overline\psi\, S_{\mu\nu} = -\cfrac{i}{2} \, \omega^{ab}(x)\, \overline\psi\, S_{ab}\,.\nonumber
\end{align}
This generalization simply promotes transformation \eqref{spinor_transformations_flat_spacetime} to gauge transformations.

Regular derivatives shall be replaced with covariant derivatives defined as follows:
\begin{align}\label{spinor_transformations_curved_spacetime}
  \nabla_\mu \psi = \pd_\mu \psi -\cfrac{i}{2}\, \left(\Gamma_\mu\right)^{ab}\, S_{ab}\, \psi \, ,\\
  \nabla_\mu \overline\psi = \pd_\mu \overline\psi +\cfrac{i}{2}\, \left(\Gamma_\mu\right)^{ab}\,\overline\psi\, S_{ab} \, . \nonumber
\end{align}
Covariant derivatives defined in such a way satisfy the following relations:
\begin{align}
  \nabla_\mu (\,\overline\psi\, \psi ) &= \pd_\mu (\,\overline\psi\, \psi)  \, & \nabla_\mu \left(\,\overline\psi\, \gamma^\mu \psi \right) &= \pd_\mu \left(\, \overline\psi\, \gamma^\mu \psi\right) + \Gamma^\alpha_{\mu\beta} \, \overline\psi\, \gamma^\beta \,\psi \,.
\end{align}
This construction produces the following generalization of the Dirac action
\begin{align}
  \mathcal{A}_{s=\frac{1}{2},m\not=0}=&\int d^4 x \sqrt{-g} \, \left[\cfrac{i}{2} \, \left(\,\overline\psi \,\gamma^\mu \nabla_\mu \psi - \nabla_\mu \overline\psi \,\gamma^\mu \psi\right) - m \,\overline\psi\,\psi \right] .\label{Dirac_action}
\end{align}
The rest of this section is devoted to a perturbative expansion of this action.

The following theorem express the Dirac action in terms of spin-connection.
\begin{theorem}
  \begin{align}
    \mathcal{A}_{s=\frac{1}{2},m\not=0}=&\int d^4 x \sqrt{-g} \, \left[\cfrac{i}{2} \, \left(\,\overline\psi \,\gamma^\mu \pd_\mu \psi - \pd_\mu \overline\psi \,\gamma^\mu \psi\right) - m \,\overline\psi\,\psi + \cfrac14\left(\Gamma_\mu\right)^{ab} \overline\psi\{\gamma^\mu,S_{ab}\}\psi \right] \, 
  \end{align}
\end{theorem}
\noindent To proof this theorem one shall use the definition of covariant derivatives \eqref{spinor_transformations_curved_spacetime}. This expression can be simplified even further if definitions of $S_{ab}$ and spin-connection \eqref{spin-connection} are used.

\begin{theorem}
  \begin{align}
    \mathcal{A}_{s=\frac{1}{2},m\not=0}=&\int d^4 x \sqrt{-g} \, \left[\cfrac{i}{2} \, \left(\,\overline\psi \,\gamma^\mu \pd_\mu \psi - \pd_\mu \overline\psi \,\gamma^\mu \psi\right) - m \,\overline\psi\,\psi  \right] \\
    &+ \int d^4 x \sqrt{-g} \, \overline\psi (\gamma^b \gamma^m\gamma^a - \gamma^a \gamma^m \gamma^b) \psi ~\cfrac{i}{8} \, \mathfrak{e}_m{}^\mu \,\mathfrak{e}_{a\lambda} \pd_\mu \mathfrak{e}_b{}^\lambda. \nonumber
  \end{align}
\end{theorem}
\begin{proof}
  First of all, the definition of $S_{ab}$ should be used:
  \begin{align}
    \cfrac14\,\{\gamma^m, S^{ab}\} = \cfrac{i}{16}\, \{\gamma^m,[\gamma^a,\gamma^b]\} = \cfrac{i}{8} \,\left( \gamma^b \gamma^m\gamma^a - \gamma^a \gamma^m \gamma^b\right).
  \end{align}
  Therefore the part containing spin-connections reads
  \begin{align}
    \overline\psi (\gamma^b\gamma^m\gamma^a-\gamma^a\gamma^m\gamma^b)\psi \,\cfrac{i}{8} \,\mathfrak{e}_m{}^\mu \left(\Gamma_\mu\right)_{ab}.
  \end{align}
  Next, the definition of spin-connection \eqref{spin-connection} should be used:
  \begin{align}
    &\overline\psi (\gamma^b\gamma^m\gamma^a-\gamma^a\gamma^m\gamma^b)\psi \,\cfrac{i}{8} \,\mathfrak{e}_m{}^\mu \left(\Gamma_\mu\right)_{ab} \\
    &= \overline\psi (\gamma^b\gamma^m\gamma^a-\gamma^a\gamma^m\gamma^b)\psi \left[\cfrac{i}{16}\,\mathfrak{e}_m{}^\mu\mathfrak{e}_a{}^\alpha\mathfrak{e}_b{}^\beta\, \left(\pd_\mu g_{\beta\alpha} + \pd_\beta g_{\mu\alpha} - \pd_\alpha g_{\mu\beta}\right) + \cfrac{i}{8}\,\mathfrak{e}_{a\lambda}\pd_\mu \mathfrak{e}_b{}^\lambda\right]\nonumber.
  \end{align}
  The part of this expression containing $\gamma$ matrices is anti-symmetric with respect to indices $a$ and $b$. Consequently, the term containing $\pd_\mu g_{\beta\alpha}$ vanishes and the expressions simplifies:
  \begin{align}
    \overline\psi (\gamma^b\gamma^m\gamma^a-\gamma^a\gamma^m\gamma^b)\psi \left[\cfrac{i}{16}\,\mathfrak{e}_m{}^\mu\mathfrak{e}_a{}^\alpha\mathfrak{e}_b{}^\beta\, \left(\pd_\beta g_{\mu\alpha} - \pd_\alpha g_{\mu\beta}\right) + \cfrac{i}{8}\,\mathfrak{e}_{a\lambda}\pd_\mu \mathfrak{e}_b{}^\lambda\right].
  \end{align}
  The last part of this expression cannot be canceled, but the first the terms vanish due to the symmetry:
  \begin{align}
    &(\gamma^b\gamma^m\gamma^a-\gamma^a\gamma^m\gamma^b) \, \mathfrak{e}_m{}^\mu\mathfrak{e}_a{}^\alpha\mathfrak{e}_b{}^\beta\, \left(\pd_\beta g_{\mu\alpha} - \pd_\alpha g_{\mu\beta}\right)= (\gamma^\beta\gamma^\mu\gamma^\alpha-\gamma^\alpha\gamma^\mu\gamma^\beta)\left(\pd_\beta g_{\mu\alpha} - \pd_\alpha g_{\mu\beta}\right)\\
    &=\gamma^\beta\gamma^\mu\gamma^\alpha \pd_\beta g_{\mu\alpha}-\gamma^\alpha\gamma^\mu\gamma^\beta \pd_\beta g_{\mu\alpha} -\gamma^\beta\gamma^\mu\gamma^\alpha \pd_\alpha g_{\mu\beta}+\gamma^\alpha\gamma^\mu\gamma^\beta \pd_\alpha g_{\mu\beta} \nonumber\\
    & = 2\, \pd_\mu  g_{\alpha\beta} \left(\gamma^\mu\gamma^\alpha\gamma^\beta - \gamma^\alpha\gamma^\beta\gamma^\mu\right) = 2\, \pd_\mu  g_{\alpha\beta} \left( (2\,g^{\mu\alpha} -\gamma^\alpha\gamma^\mu)\gamma^\beta - \gamma^\alpha(2 \,g^{\mu\beta} - \gamma^\mu\gamma^\beta) \right)\nonumber\\
    &=4\, \pd_\mu g_{\alpha\beta} ( g^{\mu\alpha} \gamma^\beta - g^{\mu\beta}\gamma^\alpha) =0 .\nonumber
  \end{align}
  Which completes the proof.
\end{proof}

Let us highlight the fact that we did not use any perturbative expansions within this proof. In turn, within the perturbative treatment the discussed action will be simplified even further.
Firstly, let us specify the perturbative expansion of vierbeins by the following theorem. The theorem is proven in \cite{Prinz:2020nru}, so we present it without a prove.
\begin{theorem}
  \begin{align}
    \mathfrak{e}_m{}^\mu & = \sum\limits_{n=0}^\infty \kappa^n \, \binom{-\frac12}{n} \, \left(h^n\right)_m{}^\mu  \, ,& \mathfrak{e}^m{}_\mu &= \sum\limits_{n=0}^\infty \kappa^n \,\binom{\frac12}{n} \, \left(h^n\right)^m{}_\mu \,.
  \end{align}
  Here $\binom{x}{y}$ are binomial coefficients.
\end{theorem}
\begin{corollary}
  \begin{align}
    \mathfrak{e}_{m\mu} = \eta_{mn} \mathfrak{e}^n{}_\mu = \sum\limits_{n=0}^\infty \kappa^n \binom{\frac12}{n} \left( h^n\right)_{m\mu}.
  \end{align}
\end{corollary}

These perturbative expansions provide a way to proof the following theorem.
\begin{theorem}
  On perturbative expansion \eqref{The_perturbative_expansion}
  \begin{align}
    (\gamma^b\gamma^m\gamma^a - \gamma^a \gamma^m\gamma^b) \mathfrak{e}_{a\lambda}\pd_\mu \mathfrak{e}_b{}^\lambda =0.
  \end{align}
\end{theorem}

\begin{proof}

  The core of the proof is the fact that only symmetric matrices enter the vierbein perturbative expansion. Let us show this in details. Firstly, the $n$-th perturbation order of $\mathfrak{e}_{a\lambda}$ within the Fourier representation reads:
  \begin{align}
    \left(\mathfrak{e}_{a\lambda}\right)^{(n)} \to \kappa^n \binom{\frac12}{n} \,(h^n)_{a\lambda} = \kappa^n \binom{\frac12}{n} h_{a}{}^\sigma(k_1)\cdots h_{\rho\lambda}(k_n).
  \end{align}
  The $n$-th perturbative order of $\pd_\mu \mathfrak{e}_b{}^\lambda$ has a similar form:
  \begin{align}
    \left(\pd_\mu \mathfrak{e}_b{}^\lambda\right)^{(n)} \to i\,\sum\limits_{p=1}^n \left\{k_p\right\}\,\kappa^n \,\binom{-\frac12}{n} h_b{}^\sigma(k_1)\cdots h_\rho{}^\lambda (k_n)\,.
  \end{align}
  Therefore a multiplication of $n_1$-th order term of $\mathfrak{e}_{a\lambda}$ on an $n_2$-th order term from $\pd_\mu \mathfrak{e}_b{}^\lambda$ results in the following expression:
  \begin{align}
    \left(\mathfrak{e}_{a\lambda}\right)^{(n_1)} \left(\pd_\mu \mathfrak{e}_b{}^\lambda\right)^{(n_2)} \to  \kappa^{n_1+n_2} \left( i\,\sum\limits_{p=1}^{n_2} \left\{k_p\right\}\right) \binom{\frac12}{n_1} \binom{-\frac12}{n_2} h_{a}{}^{\sigma}(k_1)\cdots h_{\rho\lambda}(k_{n_1}) h_b{}^{\sigma'}(k_1)\cdots h_{\rho'}{}^\lambda (k_n).
  \end{align}
  This expression contains a multiplication of $n_1+n_2$ symmetric matrices $h_{\mu\nu}$, therefore the expression is symmetric with respect to $a\leftrightarrow b$. Consequently, its contraction with an anti-symmetric expression vanishes.
\end{proof}

\begin{corollary}
  On perturbative expansion \eqref{The_perturbative_expansion}
  \begin{align}\label{Dirac_action_perturbative}
    \mathcal{A}_{s=\frac{1}{2},m\not=0}=&\int d^4 x \sqrt{-g} \, \left[\cfrac{i}{2}\,\mathfrak{e}_m^\mu \, \left(\,\overline\psi \,\gamma^m \pd_\mu \psi - \pd_\mu \overline\psi \,\gamma^m \psi\right) - m \,\overline\psi\,\psi  \right] .
  \end{align}
\end{corollary}

The perturbative structure of the Dirac action \eqref{Dirac_action_perturbative} is given by the following theorem.
\begin{theorem}
  \begin{align}
    \mathcal{A}_{s=\frac{1}{2},m\not=0} =& \sum\limits_{n=0}^\infty \int \cfrac{d^4 p_1}{(2\pi)^4}\cfrac{d^4 p_2}{(2\pi)^4} \prod\limits_{i=1}^n \cfrac{d^4 k_i}{(2\pi)^4} \, (2\pi)^4 \delta\left(p_1 + p_2 +\sum\limits_i k_i\right)\\
    &\times\kappa^n \,\overline\psi(p_2) \left[ \cfrac12\,\gamma^m(p_2-p_1)_\mu \,\mathcal{C}_E{}_m{}^{\mu\rho_1\sigma_1\cdots\rho_n\sigma_n} - m\, \mathcal{C}^{\rho_1\sigma_1\cdots\rho_n\sigma_n} \right] \psi(p_1). \nonumber
  \end{align}
  Here
  \begin{align}
    \mathcal{C}_E{}_m{}^{\mu\rho_1\sigma_1\cdots\rho_n\sigma_n} =\operatorname{Symm} \mathop{\sum\limits_{n_0,n_1=0}^\infty}_{n_0+n_1=n} \,\binom{-\frac12}{n_1} \left( \mathcal{C}_{(n_0)}{}^{\cdots} \, I_{(1+n_1)}{}_m{}^{\mu\cdots}\right)^{\rho_1\sigma_1\cdots\rho_n\sigma_n}
  \end{align}
  and the same notations are used.
\end{theorem}

The resulted expression does not require any additional symmetrization and can be used to obtain the corresponding Feynman rules. We shall also note, that the factor ``2'' shall not be introduced here as there is only one way to couple this vertex to fermion lines.

\begin{theorem}
  Expressions $\mathfrak{V}^{(n)}_{s=\frac12,m\not=0}$ that corresponds to a vertex describing interaction of two Dirac fermions and $n$ gravitons in given by the following formula.
  \begin{align}
    \left(\mathfrak{V}_{s=\frac12,m\not=0}^{(n)}\right){}^{\rho_1\sigma_1\cdots\rho_n\sigma_n}(p_1,p_2)=i\, \kappa^n\, \left[\cfrac12\,\gamma^m(p_2-p_1)_\mu \,\mathcal{C}_E{}_m{}^{\mu\rho_1\sigma_1\cdots\rho_n\sigma_n} - m\, \mathcal{C}^{\rho_1\sigma_1\cdots\rho_n\sigma_n}\right] .
  \end{align}
\end{theorem}

\section{FeynGrav}\label{Section_Implementation}

The framework described in previous sections is implemented in FeynGrav package which works as an extension of FeynCalc. Its source code is publicly available \cite{FeynGrav}. The package requires FeynCalc version 9.3.1 (or higher). Installation instructions and descriptions of files are given in ``README'' file. Finally, file ``FeynGrav\_Examples.nb'' contains annotated calculations presented in this paper.

Let us also explain the motivation to use FeynCalc and to comment on xAct bundle that can be used in further development of FeynGrav. FeynCalc is, perhaps, the most natural software to implement the presented algorithm for Feynman rules calculation. Firstly, it contains a wast array of tools that can be implemented for calculations of various matrix elements. Secondly, it contains expressions for Feynman rules for quantum chromodynamics. Lastly, it contains a tool to operate with one-loop Passarino-Veltman integrals \cite{Shtabovenko:2016sxi,Shtabovenko:2020gxv,Mertig:1990an,Passarino:1978jh}. Consequently, FeynCalc contains all the tools required for an analysis of perturbative quantum gravity.

The xAct bundle \cite{xActBundle,Portugal:1998qi,MARTINGARCIA2008597,Brizuela:2008ra,Pitrou:2013hga,Nutma:2013zea} is related with a problem similar to the calculation of Feynman rules. Namely, xPert package \cite{Brizuela:2008ra} provides a tool to evaluate perturbative expansions, including the perturbative expansion of the Einstein-Hilbert action \eqref{Hilbert_action}. One can naturally expect to find an opportunity to use a perturbative expansion generated by the xPert to derive the corresponding Feynman rules. However, we believe that a direct usage of the xPert may not result in an immediate efficiency improvement. The explicit form of an interaction Lagrangian is not as important as the Lorentz structure of the Lagrangian, that is the way to contract indices between fields $h_{\mu\nu}$ and their momenta. The formalism presented in this paper provides an explicit way to factorize that structure. On the contrary, xPert only provides a user with a symbolic expression, so the structure of an interaction term remains implicit. Still, the opportunity to use the xAct bundle for Feynman rules calculation must not be discarded. In particular, package xPerm \cite{MARTINGARCIA2008597} are expected to provide a much productive framework to operate with tensor indices. This opportunity will be explored in further publications.

In this paper we will not discuss the source code in details because of two reasons. Firstly, the code does not rely on any sophisticated techniques that worth to be discussed. The most part of calculations are based on manipulations with arrays of Lorentz indices which may look massive, but do not require any sophisticated techniques. Secondly, we would like to focus on results that FeynGrav helps to obtain.

We would only like to give the following comment on the technical side of the package. First and foremost, all expressions for interaction vertices used in the package are taken from pre-made libraries. By default these libraries contain data for perturbation theory up to $\mathcal{O}(\kappa^3)$ order. This is done for the sake of performance optimization as the corresponding expressions take quite some time to evaluate. Expressions for vertices can also be generated independently with ``Libraries\_Generator.nb'' file. Secondly, $I$ tensors are evaluate in exactly the same way as described in the previous section. The script generates a single term and then performs a consecutive symmetrization. $C$ tensors are evaluated in a similar way. Namely, the sum of $I$ tensors is evaluated and the result is symmetrized. Tensors $\mathcal{C}_{I}$, $\mathcal{C}_{II}$, $\mathcal{C}_{III}$, and $\mathcal{C}_{E}$ are evaluated in the same way, i.e. the script performs a summation and then performs a symmetrization. Finally, expressions for interaction vertices are evaluated similarly. The script obtains an expression for a single pair of momenta and then summarizes over all required pairs of momenta.

The full list of FeynGrav commands is given in Appendix \ref{appendix_commands} and can also be called within the program itself with ``FeynGravCommands''. There are three classes of commands. Firstly, there are commands that return expressions for interaction vertices. Secondly, there are commands that return expressions for the graviton propagator. The propagator (and all other expressions) is evaluated in the harmonic gauge which is fixed by the following term:
\begin{align}
  \mathcal{L}_\text{gf} = \int d^4 x \, \left( \pd_\mu h^{\mu\nu} - \cfrac12\, \pd^\nu h \right)^2\,.
\end{align}
The resulting expression for the graviton propagator reads:
\begin{align}\label{Graviton_propagator}
  \mathcal{G}_{\mu\nu\alpha\beta}(k) = i \, \cfrac{\cfrac12\left[ \eta_{\mu\alpha}\eta_{\nu\beta} + \eta_{\mu\beta}\eta_{\nu\alpha} - \eta_{\mu\nu} \eta_{\alpha\beta}\right]}{k^2}\,.
\end{align}
There are three commands related with the graviton propagator. They are introduced as a mean to simplify some calculations. Their description is also given in Appendix \ref{appendix_commands}.

\subsection{$2\to 2$ on-shell graviton scattering}\label{graviton_scattering_example}

Let us turn to a discussion of particular calculations where FeynGrav can be implemented. Perhaps the best example of FeynGrav usage is provided by $2 \to 2$ tree-level on-shell graviton scattering. For the best of our knowledge, the corresponding amplitudes were first calculated in \cite{Sannan:1986tz}. We will closely follow the calculations presented in \cite{Sannan:1986tz} to make the result easily comparable.

In order to calculate on-shell scattering amplitudes graviton polarization operators should be defined. Graviton polarization tensors $\varepsilon_{\mu\nu}(p)$ are defined via the standard polarization vectors $\varepsilon_{\mu}(p)$:
\begin{align}
  \varepsilon_{\mu\nu} (p) = \varepsilon_\mu (p) \varepsilon_\nu (p)\,.
\end{align}
The kinematics of the process is discussed in \cite{Sannan:1986tz} in details, so we will adopt it directly from that paper.
The two following issues must be noted.
Firstly, in \cite{Sannan:1986tz} metric signature $(-+++)$ was used, which results in a different overall sing of the amplitudes.
Secondly, it order to obtain an answer easily comparable with \cite{Sannan:1986tz} we adopt chirality notations from \cite{Sannan:1986tz}. These notations, however, differ from notations used in contemporary literature, which may cause some confusion. Still, we adopt notations from \cite{Sannan:1986tz} for the sake of comparison.

FeynCalc provides a realization of polarization operators by means of ``PolarizationVector'' command which refers to the following object:
\begin{lstlisting}[language=Mathematica]
  Pair[LorentzIndex[\[Mu]], Momentum[Polarization[p, I]]] 
\end{lstlisting}
The conjugated polarization vector is defined with the following command:
\begin{lstlisting}[language=Mathematica]
  Pair[LorentzIndex[\[Mu]], Momentum[Polarization[p, -I]]]
\end{lstlisting}

The complete code evaluating the discussed amplitude can be found in the example file in \cite{FeynGrav}. The $s$-channel amplitude for $2\to 2$ tree-level graviton scattering is given by the following expression:
\begin{align}
  \begin{gathered}
    \begin{fmffile}{D07}
      \begin{fmfgraph}(30,30)
        \fmfbottom{B1,B2}
        \fmftop{T1,T2}
        \fmf{dbl_wiggly}{B1,D,B2}
        \fmf{dbl_wiggly}{T1,U,T2}
        \fmf{dbl_wiggly}{D,U}
        \fmfdot{U,D}
      \end{fmfgraph}
    \end{fmffile}
  \end{gathered}
  = \mathcal{M}_s=&\,\text{GravitonVertex}[\mu_1,\nu_1,p_1,\mu_2,\nu_2,p_2,\alpha_1,\beta_1,-p_1-p_2] \nonumber \\
  & \times\text{GravitonPropagatorAlternative}[\alpha_1,\beta_1,\alpha_2,\beta_2,p_1 + p_2] \nonumber \\
  & \times\text{GravitonVertex}[\mu_3,\nu_3,p_3,\mu_4,\nu_4,p_4,\alpha_2,\beta_2, p_1 + p_2]\nonumber
\end{align}
Amplitudes for t- and u-channels are obtained from the s-amplitude via cross-symmetry. The amplitude corresponding to the four-graviton interaction reads:
\begin{align}
  \begin{gathered}
    \begin{fmffile}{D08}
      \begin{fmfgraph}(30,30)
        \fmfbottom{B1,B2}
        \fmftop{T1,T2}
        \fmf{dbl_wiggly}{B1,V,B2}
        \fmf{dbl_wiggly}{T1,V,T2}
        \fmf{dbl_wiggly}{V}
        \fmfdot{V}
      \end{fmfgraph}
    \end{fmffile}
  \end{gathered}
  =\mathcal{M}_4 =\text{GravitonVertex}[\mu_1,\nu_1,p_1,\mu_2,\nu_2,p_2,\mu_3,\nu_3,p_3,\mu_4,\nu_4,p_4].
\end{align}
The calculated amplitudes takes the following form
\begin{align}
  \mathcal{M}(++++)&=i\,\cfrac{\kappa^2}{4}\,\cfrac{s^4}{s\,t\,u}\,,&  \mathcal{M}(+-+-)&=i\,\cfrac{\kappa^2}{4}\,\cfrac{u^4}{s\,t\,u}\,,&  \mathcal{M}(+--+)&=i\,\cfrac{\kappa^2}{4}\,\cfrac{t^4}{s\,t\,u}\,,&
\end{align}
\begin{align*}
  \mathcal{M}(++--)=\mathcal{M}(+++-)=0\,.
\end{align*}
These expression matches expressions given in \cite{Sannan:1986tz} up to the overall sign, which, as was noted above, is due to the difference in the choice of the metric signature. Let us also note one more time that we used the chirality notations from \cite{Sannan:1986tz} which does not match the contemporary notations which can be found, for instance, in \cite{Elvang:2013cua}.

Finally, for the sake of generality we can present an expression for $2\to 2$ graviton on-shell scattering in arbitrary $d$ in kinematics taken from \cite{Sannan:1986tz}:
\begin{align}
  \mathcal{M}(h_1,h_2,h_3,h_4) =& \cfrac{i\,\kappa^2}{128}\,\cfrac{1}{s\,t\,u} \, \cfrac{1}{s^2} \Big[ 8 h_2 h_3 t^6+8 h_1 h_4 t^6+8 t^6+6 d^2 h_1 h_3 u t^5-36 d h_1 h_3 u t^5+88 h_1 h_3 u t^5\\
    &+32 h_2 h_3 u t^5+32 h_1 h_4 u t^5+6 d^2 h_2 h_4 u t^5-36 d h_2 h_4 u t^5+88 h_2 h_4 u t^5+32 u t^5\nonumber \\
    &+8 h_1 h_2 h_3 h_4 s^2 t^4+32 d^2 u^2 t^4-192 d u^2 t^4-2 d^2 h_1 h_3 u^2 t^4+12 d h_1 h_3 u^2 t^4+80 h_1 h_3 u^2 t^4\nonumber \\
    & +56 h_2 h_3 u^2 t^4+56 h_1 h_4 u^2 t^4-2 d^2 h_2 h_4 u^2 t^4+12 d h_2 h_4 u^2 t^4+80 h_2 h_4 u^2 t^4+320 u^2 t^4\nonumber \\
    &+44 d^2 u^3 t^3-264 d u^3 t^3+88 h_1 h_3 u^3 t^3+88 h_2 h_3 u^3 t^3+88 h_1 h_4 u^3 t^3+88 h_2 h_4 u^3 t^3\nonumber \\
    &+432 u^3 t^3+88 h_1 h_2 s^2 u t^3+3 d^2 h_1 h_3 s^2 u t^3-18 d h_1 h_3 s^2 u t^3+3 d^2 h_2 h_4 s^2 u t^3-18 d h_2 h_4 s^2 u t^3\nonumber\\
    &+16 h_1 h_2 h_3 h_4 s^2 u t^3+88 h_3 h_4 s^2 u t^3+32 d^2 u^4 t^2-192 d u^4 t^2+56 h_1 h_3 u^4 t^2-2 d^2 h_2 h_3 u^4 t^2\nonumber \\
    & +12 d h_2 h_3 u^4 t^2+80 h_2 h_3 u^4 t^2-2 d^2 h_1 h_4 u^4 t^2+12 d h_1 h_4 u^4 t^2+80 h_1 h_4 u^4 t^2+56 h_2 h_4 u^4 t^2\nonumber \\
    &+320 u^4 t^2+168 h_1 h_2 s^2 u^2 t^2-d^2 h_1 h_3 s^2 u^2 t^2+6 d h_1 h_3 s^2 u^2 t^2-d^2 h_2 h_3 s^2 u^2 t^2 \nonumber \\
    &+6 d h_2 h_3 s^2 u^2 t^2-d^2 h_1 h_4 s^2 u^2 t^2 +6 d h_1 h_4 s^2 u^2 t^2-d^2 h_2 h_4 s^2 u^2 t^2 + 6 d h_2 h_4 s^2 u^2 t^2 \nonumber \\
    &+32 d^2 h_1 h_2 h_3 h_4 s^2 u^2 t^2-192 d h_1 h_2 h_3 h_4 s^2 u^2 t^2+280 h_1 h_2 h_3 h_4 s^2 u^2 t^2+168 h_3 h_4 s^2 u^2 t^2\nonumber \\
    &+32 h_1 h_3 u^5 t+6 d^2 h_2 h_3 u^5 t-36 d h_2 h_3 u^5 t+88 h_2 h_3 u^5 t+6 d^2 h_1 h_4 u^5 t-36 d h_1 h_4  u^5 t\nonumber \\
    & +88 h_1 h_4 u^5 t+32 h_2 h_4 u^5 t+32 u^5 t+88 h_1 h_2 s^2 u^3 t+3 d^2 h_2 h_3 s^2 u^3 t-18 d h_2 h_3 s^2 u^3 t\nonumber \\
    &+3 d^2 h_1 h_4 s^2 u^3 t-18 d h_1 h_4 s^2 u^3 t+16 h_1 h_2 h_3 h_4 s^2 u^3 t+88 h_3 h_4 s^2 u^3 t+9 d^2 h_1 h_2 s^4 u t\nonumber \\
    &-54 d h_1 h_2 s^4 u t+9 d^2 h_3 h_4 s^4 u t-54 d h_3 h_4 s^4 u t+8 h_1 h_3 u^6+8 h_2 h_4 u^6+8 u^6+8 h_1 h_2 h_3 h_4 s^2 u^4 \Big] \,.\nonumber
\end{align}

\subsection{Polarization operators}\label{polarization_operators}

Another critical test for the applicability of FeynGrav is provided by the structure of divergencies at the one-loop level. It is well-known that general relativity without matter is UV-finite on-shell at the one-loop level \cite{tHooft:1974toh}. This can be tested via a calculation of the one-loop graviton polarization operators. Contribution of a graviton loop to the graviton polarization operator off-shell is given by the following expression:
\begin{align}
  \begin{gathered}
    \begin{fmffile}{G2G01}
        \begin{fmfgraph*}(40,40)
          \fmfleft{L}
          \fmfright{R}
          \fmf{dbl_wiggly,tension = 2}{L,VL}
          \fmf{dbl_wiggly,tension = 2}{VR,R}
          \fmf{dbl_wiggly,right=1,tension=.2}{VL,VR,VL}
          \fmfdot{VL,VR}
          \fmflabel{$\mu\nu$}{L}
          \fmflabel{$\alpha\beta$}{R}
        \end{fmfgraph*}
    \end{fmffile}
  \end{gathered}\hspace{20pt} =& \text{GravitonVertex}[\mu,\nu,p,\alpha_1, \beta_1, -k, \rho_1,\sigma_1, -(p - k)] \text{GravitonPropagator}[\alpha_1, \beta_1, \alpha_2, \beta_2, k]\\
  &\times\text{GravitonPropagator}[\rho_1, \sigma_1, \rho_2, \sigma_2, p - k] \text{GravitonVertex}[\alpha_2, \beta_2,  k, \rho_2, \sigma_2, p - k,\alpha,\beta, -p]\nonumber
\end{align}
\begin{align}
  = i \pi^2 B_0 (p^2,0,0) \,\cfrac{\kappa^2}{128}\,p^4 \Bigg[&-\cfrac{4 \left(d^4-10 d^3-137 d^2+458 d+176\right)}{d^2-1}\,P^2_{\mu\nu\alpha\beta} \\
    &+ \cfrac{2 \left(3 d^6-60 d^5+433 d^4-808 d^3-1172 d^2+1988 d+3896\right)}{d^2-1} \, P^0_{\mu\nu\alpha\beta}\nonumber \\
    &+\cfrac{16 (d-2)(3d-11)}{d-1}\,P^1_{\mu\nu\alpha\beta} + 8 (d-2)(d-3)\,\overline{P^0}_{\mu\nu\alpha\beta} \nonumber \\
    & +\cfrac{4 (d-2)\left(d^3-12 d^2+44 d-46\right)}{d-1}\, \overline{\overline{P^0}}_{\mu\nu\alpha\beta} \Bigg]\,. \nonumber
\end{align}
Here $B_0$ is a Passarino-Veltman integral \cite{Mertig:1990an} and $P^2$, $P^0$, $P^1$, $\overline{P^0}$, $\overline{\overline{P^0}}$ are Nieuwenhuizen operators \cite{VanNieuwenhuizen:1973fi,Accioly:2000nm,Latosh:2020ysu}. When the momenta $p$ is fixed on-shell the expression simplifies:
\begin{align}
  \begin{gathered}
    \begin{fmffile}{G2G01}
        \begin{fmfgraph*}(40,40)
          \fmfleft{L}
          \fmfright{R}
          \fmf{dbl_wiggly,tension = 2}{L,VL}
          \fmf{dbl_wiggly,tension = 2}{VR,R}
          \fmf{dbl_wiggly,right=1,tension=.2}{VL,VR,VL}
          \fmfdot{VL,VR}
          \fmflabel{$\mu\nu$}{L}
          \fmflabel{$\alpha\beta$}{R}
        \end{fmfgraph*}
    \end{fmffile}
  \end{gathered} \hspace{1cm}= (d-2)(d^3-18 d^2+108 d-192)\,\cfrac{\kappa^2}{16} \, p_\mu p_\nu p_\alpha p_\beta \, \int \cfrac{d^4k}{(2\pi)^4} \, \cfrac{1}{k^2 (k-p)^2}\,.
\end{align}
In order to fix the expression on-shell completely one should (after a few intermediate steps) multiply the expression on the corresponding polarization operators $\varepsilon_{\mu\nu}(p)$ which makes it vanish due to the gauge symmetry. Therefore, in full agreement with the known results, pure gravity at the one-loop level remains UV-finite.

Contributions to the graviton polarization operator from massless matter loops evaluated off-shell are presented below.
\begin{align}
  \begin{gathered}
    \begin{fmffile}{G2G02}
        \begin{fmfgraph*}(40,40)
          \fmfleft{L}
          \fmfright{R}
          \fmf{dbl_wiggly,tension = 2}{L,VL}
          \fmf{dbl_wiggly,tension = 2}{VR,R}
          \fmf{dashes,right=1,tension=.2}{VL,VR,VL}
          \fmfdot{VL,VR}
          \fmflabel{$\mu\nu$}{L}
          \fmflabel{$\alpha\beta$}{R}
        \end{fmfgraph*}
    \end{fmffile}
  \end{gathered}\hspace{20pt} =&  i \pi^2 B_0 (p^2,0,0) \,\cfrac{\kappa^2}{16}\,\cfrac{1}{d^2-1}\,p^4 \Bigg[2\,P^2_{\mu\nu\alpha\beta} +  \left(3 d^2-6 d-4\right) \, P^0_{\mu\nu\alpha\beta}\Bigg]\,;\\
  \begin{gathered}
    \begin{fmffile}{G2G03}
        \begin{fmfgraph*}(40,40)
          \fmfleft{L}
          \fmfright{R}
          \fmf{dbl_wiggly,tension = 2}{L,VL}
          \fmf{dbl_wiggly,tension = 2}{VR,R}
          \fmf{photon,right=1,tension=.2}{VL,VR,VL}
          \fmfdot{VL,VR}
          \fmflabel{$\mu\nu$}{L}
          \fmflabel{$\alpha\beta$}{R}
        \end{fmfgraph*}
    \end{fmffile}
  \end{gathered}\hspace{20pt}=&  i \pi^2 B_0 (p^2,0,0) \,\cfrac{\kappa^2}{16}\,\cfrac{1}{d^2-1}\,p^4 \Bigg[2 \left(2 d^2-3 d-8\right)\,P^2_{\mu\nu\alpha\beta} + (d-4)(3d^2-8d-8) \, P^0_{\mu\nu\alpha\beta}\Bigg]\,;\nonumber\\
  \begin{gathered}
    \begin{fmffile}{G2G04}
        \begin{fmfgraph*}(40,40)
          \fmfleft{L}
          \fmfright{R}
          \fmf{dbl_wiggly,tension = 2}{L,VL}
          \fmf{dbl_wiggly,tension = 2}{VR,R}
          \fmf{fermion,right=1,tension=.2}{VL,VR,VL}
          \fmfdot{VL,VR}
          \fmflabel{$\mu\nu$}{L}
          \fmflabel{$\alpha\beta$}{R}
        \end{fmfgraph*}
    \end{fmffile}
  \end{gathered}\hspace{20pt} =& i \pi^2 B_0 (p^2,0,0) \,\cfrac{\kappa^2}{16}\,p^4 \, \cfrac{2-d}{d-1} \, P^1_{\mu\nu\alpha\beta}\,.\nonumber
\end{align}
Similarly to the previous case these polarization operators vanish if $p$ is fixed on-shell. This result takes place only due to the dimensional reasons. At the one-loop level graviton polarization operator is proportional to $\kappa^2$ which has mass dimension $-2$. If a matter field is massless, the only dimensional parameter at hand is $p^2$. As soon as it is fixed on-shell it becomes impossible to compensate the negative dimension of $\kappa$ and the polarization operator can only vanish. Therefore the corresponding matrix elements remain UV-finite on-shell even if massless matter is added to gravity.

In a similar manner one-loop gravitational corrections for massless scalar, vector, and fermion propagators can be calculated. The corresponding expressions are given below.
\begin{align}
  \begin{gathered}
    \begin{fmffile}{P1}
      \begin{fmfgraph}(40,40)
        \fmfleft{L}
        \fmfright{R}
        \fmf{dashes}{L,VL,VR,R}
        \fmf{dbl_wiggly,left=1,tension=.2}{VL,VR}
        \fmfdot{VL,VR}
      \end{fmfgraph}
    \end{fmffile}
  \end{gathered}
  =&i\pi^2\,B_0(p^2,0,0) \, \cfrac{\kappa^2}{32} \,(d-4)(2-d)\,p^4\, ;\\
  \begin{gathered}
    \begin{fmffile}{P2}
      \begin{fmfgraph*}(40,40)
        \fmfleft{L}
        \fmfright{R}
        \fmf{photon}{L,VL,VR,R}
        \fmf{dbl_wiggly,left=1,tension=.2}{VL,VR}
        \fmflabel{$\lambda_1$}{L}
        \fmflabel{$\lambda_2$}{R}
        \fmfdot{VL,VR}
      \end{fmfgraph*}
    \end{fmffile}
  \end{gathered}
  \hspace{20pt}
  =& i\pi^2 B_0 (p^2,0,0)\,\cfrac{\kappa^2}{32}\,\cfrac{(d-2)(d^2-14 d +32)}{d-1} \, p^4 \, \theta_{\lambda_1\lambda_2}(p) \, ; \nonumber\\
  \begin{gathered}
    \begin{fmffile}{P3}
      \begin{fmfgraph}(40,40)
        \fmfleft{L}
        \fmfright{R}
        \fmf{fermion}{L,VL,VR,R}
        \fmf{dbl_wiggly,left=1,tension=.1}{VL,VR}
        \fmfdot{VL,VR}
      \end{fmfgraph}
    \end{fmffile}
  \end{gathered}
  =& i\pi^2 \, B_0 (p^2,0,0) \,\cfrac{\kappa^2}{128}\, (3-2d)(2-d) \, p^2\, \gamma\cdot p \,.\nonumber
\end{align}
Here $\theta_{\mu\nu}(p) = \eta_{\mu\nu} - p_\mu p_\nu / p^2$ is the standard gauge projector.

The corresponding corrections to propagators also vanish on-shell due to the same dimensional reasoning. Although it must be noted that the scalar field polarization operator also vanishes off-shell due to $(d-4)$ factor.

\subsection{One-loop scalar-graviton interaction structure}

Another important result is related with the one-loop structure of scalar-gravitational interaction. At the one-loop level the interaction is described by the following diagrams:
\begin{align}
  \begin{gathered}
    \begin{fmffile}{T00}
      \begin{fmfgraph}(40,40)
        \fmfleft{L}
        \fmfright{R1,R2}
        \fmf{dbl_wiggly}{L,V}
        \fmf{dashes}{R1,V,R2}
        \fmfv{decor.shape=circle,decor.filled=shaded,decor.size=15}{V}
      \end{fmfgraph}
    \end{fmffile}
  \end{gathered}
  =
  \begin{gathered}
    \begin{fmffile}{T01}
      \begin{fmfgraph}(40,40)
        \fmfleft{L}
        \fmfright{R1,R2}
        \fmf{phantom,tension=2}{L,VL}
        \fmf{phantom,tension=2}{R1,VR1}
        \fmf{phantom,tension=2}{R2,VR2}
        \fmf{phantom,tension=.7}{VL,VR1,VR2,VL}
        \fmfdot{VL,VR1,VR2}
        \fmffreeze
        \fmf{dbl_wiggly}{L,VL}
        \fmf{dbl_wiggly}{VR1,VR2}
        \fmf{dashes}{R1,VR1,VL,VR2,R2}
      \end{fmfgraph}
    \end{fmffile}
  \end{gathered}
  +
  \begin{gathered}
    \begin{fmffile}{T02}
      \begin{fmfgraph}(40,40)
        \fmfleft{L}
        \fmfright{R1,R2}
        \fmf{phantom,tension=2}{L,VL}
        \fmf{phantom,tension=2}{R1,VR1}
        \fmf{phantom,tension=2}{R2,VR2}
        \fmf{phantom,tension=.7}{VL,VR1,VR2,VL}
        \fmfdot{VL,VR1,VR2}
        \fmffreeze
        \fmf{dbl_wiggly}{L,VL}
        \fmf{dbl_wiggly}{VR1,VL,VR2}
        \fmf{dashes}{R1,VR1,VR2,R2}
      \end{fmfgraph}
    \end{fmffile}
  \end{gathered}
  +
  \begin{gathered}
    \begin{fmffile}{T03}
      \begin{fmfgraph}(40,40)
        \fmfleft{L}
        \fmfright{R1,R2}
        \fmf{phantom,tension=1.7}{L,VL}
        \fmf{phantom,tension=1.7}{R1,VR,R2}
        \fmf{phantom}{VL,VR}
        \fmfdot{VL,VR}
        \fmffreeze
        \fmf{dbl_wiggly}{L,VL}
        \fmf{dbl_wiggly,right=1}{VL,VR,VL}
        \fmf{dashes}{R1,VR,R2}
      \end{fmfgraph}
    \end{fmffile}
  \end{gathered}
  +
  \begin{gathered}
    \begin{fmffile}{T04}
      \begin{fmfgraph}(40,40)
        \fmfleft{L}
        \fmfright{R1,R2}
        \fmf{dbl_wiggly}{L,V}
        \fmf{dashes}{R1,V}
        \fmf{dashes}{R2,V}
        \fmffreeze
        \fmf{phantom}{R1,VR,V}
        \fmfdot{V,VR}
        \fmffreeze
        \fmf{dbl_wiggly,left=1}{V,VR}
      \end{fmfgraph}
    \end{fmffile}
  \end{gathered}
  +
  \begin{gathered}
    \begin{fmffile}{T05}
      \begin{fmfgraph}(40,40)
        \fmfleft{L}
        \fmfright{R1,R2}
        \fmf{dbl_wiggly}{L,V}
        \fmf{dashes}{R1,V}
        \fmf{dashes}{R2,V}
        \fmffreeze
        \fmf{phantom}{R2,VR,V}
        \fmfdot{V,VR}
        \fmffreeze
        \fmf{dbl_wiggly,left=1}{V,VR}
      \end{fmfgraph}
    \end{fmffile}
  \end{gathered}
  +
  \begin{gathered}
    \begin{fmffile}{T06}
      \begin{fmfgraph}(40,40)
        \fmfleft{L}
        \fmfright{R1,RR,R2}
        \fmf{dbl_wiggly}{L,V}
        \fmf{dashes}{R1,V}
        \fmf{dashes}{R2,V}
        \fmffreeze
        \fmf{dbl_wiggly,right=.8}{V,RR,V}
        \fmfdot{V}
      \end{fmfgraph}
    \end{fmffile}
  \end{gathered}
\end{align}
FeynGrav provides a tool to calculate these diagrams off-shell in an arbitrary number of dimensions. The corresponding expression is given via a set of structure functions:
\begin{align}\label{one-loop_ST_gravity}
  &(\mathcal{M}_\text{1-loop})_{\mu\nu} \\
  &= i\pi^2B_0(p_1,0,0)\Bigg[ (F_{1,1})^{B_0}_{p_1} {p_1}{}_{\mu} {p_1}{}_{\nu}\!+\!(F_{1,2})^{B_0}_{p_1} {p_2}{}_\mu {p_2}{}_\nu \!+\!(F_{2,1})^{B_0}_{p_1} {p_1}{}_\mu {p_2}{}_\nu \!+\! (F_{2,2})^{B_0}_{p_1} {p_2}{}_\mu {p_1}{}_\nu \!+\! (F_3)^{B_0}_{p_1}\eta_{\mu\nu} \Bigg] \nonumber\\
  & + i\pi^2B_0(p_2,0,0)\Bigg[ (F_{1,1})^{B_0}_{p_2} {p_1}{}_\mu {p_1}{}_\nu\!+\!(F_{1,2})^{B_0}_{p_2} {p_2}{}_\mu {p_2}{}_\nu \!+\!(F_{2,1})^{B_0}_{p_2} {p_1}{}_\mu {p_2}{}_\nu \!+\! (F_{2,2})^{B_0}_{p_2} {p_2}{}_\mu {p_1}{}_\nu \!+\! (F_3)^{B_0}_{p_2}\eta_{\mu\nu} \Bigg] \nonumber\\
  & + i\pi^2B_0(k,0,0)\Bigg[ (F_{1,1})^{B_0}_{k} {p_1}{}_\mu {p_1}{}_\nu\!+\!(F_{1,2})^{B_0}_{k} {p_2}{}_\mu {p_2}{}_\nu \!+\!(F_{2,1})^{B_0}_{k} {p_1}{}_\mu {p_2}{}_\nu \!+\! (F_{2,2})^{B_0}_{k} {p_2}{}_\mu {p_1}{}_\nu \!+\! (F_3)^{B_0}_{k}\eta_{\mu\nu} \Bigg] \nonumber\\
  & + i\pi^2C_0(p_1,p_2,k,0,0,0)\Bigg[ (F_{1,1})^{C_0} {p_1}{}_\mu {p_1}{}_\nu\!+\!(F_{1,2})^{C_0} {p_2}{}_\mu {p_2}{}_\nu \!+\!(F_{2,1})^{C_0} {p_1}{}_\mu {p_2}{}_\nu \!+\! (F_{2,2})^{C_0} {p_2}{}_\mu {p_1}{}_\nu \!+\! (F_3)^{C_0}\eta_{\mu\nu} \Bigg] \nonumber
\end{align}
Here $p_1$ and $p_2$ are scalar momenta, $k$ is the graviton momentum, and all momenta are directed inwards $k+p_1+p_2=0$. $B_0$ and $C_0$ are Passarino-Veltman integrals \cite{Mertig:1990an}; and $C_0$ is UV-finite. Finally, $(F_{1,1})^{B_0}_{p_1}$, $(F_{1,2})^{B_0}_{p_1}$, $(F_{1,1})^{B_0}_{p_2}$, etc are structure functions. We construct them in such a way so they only depend on $p_1$ and $p_2$. Some of these functions are dependent due to the $p_1 \leftrightarrow p_2$ symmetry:
\begin{align}
  (F_{1,1})^{B_0}_{p_1} &= (F_{1,2})^{B_0}_{p_2} \Big|_{p_1\leftrightarrow p_2} \,,& (F_{1,1})^{B_0}_{k} &= (F_{1,2})^{B_0}_{k} \Big|_{p_1\leftrightarrow p_2} \,, &(F_{1,1})^{C_0} &= (F_{1,2})^{C_0} \Big|_{p_1\leftrightarrow p_2} \,, \\
  (F_{1,1})^{B_0}_{p_2} &= (F_{1,2})^{B_0}_{p_1} \Big|_{p_1\leftrightarrow p_2} \,,& (F_{2,1})^{B_0}_{k} &= (F_{2,2})^{B_0}_{k} \Big|_{p_1\leftrightarrow p_2} \, , & (F_{2,1})^{C_0} &= (F_{2,2})^{C_0} \Big|_{p_1\leftrightarrow p_2} \,, \nonumber\\
  (F_{2,1})^{B_0}_{p_1} &= (F_{2,2})^{B_0}_{p_2} \Big|_{p_1\leftrightarrow p_2} \,,& (F_{3})^{B_0}_{k} &= (F_{3})^{B_0}_{k} \Big|_{p_1\leftrightarrow p_2} \, ,& (F_{3})^{C_0} &= (F_{3})^{C_0} \Big|_{p_1\leftrightarrow p_2} \,, \nonumber\\
  (F_{2,1})^{B_0}_{p_2} &= (F_{2,2})^{B_0}_{p_1} \Big|_{p_1\leftrightarrow p_2} \,, & (F_{3})^{B_0}_{p_1} &= (F_{3})^{B_0}_{p_2} \Big|_{p_1\leftrightarrow p_2} \, , & (F_{2,1})^{B_0}_{p_1} &= (F_{2,2})^{B_0}_{p_1} \,. \nonumber
\end{align}
Explicit expressions for these functions are presented in Appendix \ref{Appendix_Structure_Functions}.

It is useful to study this expression in $d=4$ with scalar field momenta begin fixed on-shell and the graviton begin off-shell. In that case the expression can describe a virtual graviton exchange, so it can be used to study interaction of a real scalar with an external gravitational field. 
\begin{align}\label{ST_interaction_on-shell}
  (\mathcal{M}_\text{1-loop})_{\mu\nu} =&\cfrac{\kappa^3}{16}\left[3\, C_{\mu\nu\alpha\beta}(p_1)^\alpha (p_2)^\beta - \cfrac23\, (p_1)_\mu (p_1)_\nu \right]\,i\,\pi^2\,B_0(p_1,0,0)\Big|_{d=4}^{p_1^2=p_2^2=0}\\
  &+\cfrac{\kappa^3}{16}\left[3\, C_{\mu\nu\alpha\beta}(p_1)^\alpha (p_2)^\beta - \cfrac23\, (p_2)_\mu (p_2)_\nu \right]\,i\,\pi^2\,B_0(p_2,0,0)\Big|_{d=4}^{p_1^2=p_2^2=0} \nonumber\\
  &+\cfrac{\kappa^3}{8}\,k^2\,\left[k^2\, \theta_{\mu\nu}(k)-\cfrac{7}{3}\,C_{\mu\nu\alpha\beta}\,(p_1)^\alpha(p_2)^\beta\right]\,i\,\pi^2\,B_0(k,0,0)\Big|_{d=4}^{p_1^2=p_2^2=0}\nonumber\\
  &-\cfrac{\kappa^3}{8}\,k^4\,C_{\mu\nu\alpha\beta}(p_1)^\alpha(p_2)^\beta\,i\,\pi^2\,C_0(0,0,k,0,0,0)\Big|_{d=4}^{p_1^2=p_2^2=0}\,. \nonumber
\end{align}
The integral $C_0$ is UV finite and it only contains IR divergencies which can be regularized in the standard way via a soft graviton radiation. In other words, the amplitude shows that a real massless scalar generates a new non-minimal UV finite interaction with gravity. This result was obtain earlier in \cite{Latosh:2020jyq,Latosh:2021usy} via a different technique. Therefore FeynGrav provides an independent evidence supporting that claim.

\section{Discussion and conclusions}\label{Section_Conclusion}

In this paper we presented a framework capable to provide a relatively simple way to operate with Feynman rules for gravity within a computer algebra system. In Section \ref{Section_tensor_framework_bosons} we presented the core of the framework. Namely, we defined $I$ and $C$ tensors that are used to obtain perturbative expansions for $g^{\mu\nu}$, $\sqrt{-g}$, $\sqrt{-g}\, g^{\mu\nu}$, $\sqrt{-g}\, g^{\mu\nu} g^{\alpha\beta}$, and $\sqrt{-g} \, g^{\mu\nu} g^{\alpha\beta} g^{\rho\sigma}$ which are typical for bosonic Lagrangians. In Sections \ref{Section_GR_rules}, \ref{Section_Scalars}, and \ref{Section_Vectors} we show how these factors shall be used to obtain Feynman rules for pure gravity, massless vectors, and scalars of an arbitrary mass. Section \ref{Section_Fermions} is completely devoted to a derivation of Feynman rules for Dirac fermions of an arbitrary mass. We briefly revisit a construction of Dirac fermions in a curved spacetime and obtained a perturbative expansion of the Dirac action.

Finally, Section \ref{Section_Implementation} is completely devoted to a discussion of ``FeynGrav'', a packaged based on FeynCalc that implements the discussed Feynman rules. Three particular examples of implementations are discussed. The first one is the tree level $2\to 2$ on-shell graviton scattering that is discussed in Section \ref{graviton_scattering_example}. The scattering amplitude was firstly evaluated in \cite{Sannan:1986tz} and analytic calculations take about a page long. Within FeynGrav the same calculations can be performed much more easily and faster. The second example is a calculation of polarization operators for graviton and matter propagators. Explicit off-shell expressions for the corresponding operators are obtained in Section \ref{polarization_operators}. All obtained polarization operators effectively vanish on-shell due to dimensional reasoning. This is in agreement with the well-known result showing that at the one-loop level pure gravity is UV finite on-shell \cite{tHooft:1974toh}. Finally, the one-loop structure of scalar-gravitational interaction was studied. The complete one-loop interaction amplitude \eqref{one-loop_ST_gravity} was calculated with FeynGrav off-shell. When scalar fields are fixed on-shell \eqref{ST_interaction_on-shell} it is shown explicitly that the interaction generates an UV finite contribution. This results is also in agreement with the previous papers \cite{Latosh:2020jyq,Latosh:2021usy}.

We believe that FeynGrav provides a suitable framework to deal with perturbative effective gravity in low orders of perturbation theory. Even in the current state the package provides a tool to calculate sophisticated off-shell amplitudes which can be used for a detailed study of one-loop gravity structure.

Finally, let us comment on further possible development of FeynGrav. There are a few way to extend the package functional. First and foremost, more sophisticated models of matter can be added. Gauge $\operatorname{SU}(\operatorname{N})$ Yang-Mills theory serves as the most obvious candidate. It appears that Feynman rules for models that do not contain non-minimal interactions with gravity can be obtained with the use of $I$ and $C$ tensors defined in this paper. Secondly, models with non-minimal interactions between matter and gravity can be implemented. This case expected to be more complicated because such non-minimal interactions admit more sophisticated structure that may not be reducible to a combination of $I$ and $C$ tensors. Thirdly, modifications of the gravitational sector can be considered. This direction of development appears to be the most sophisticated. The main reason is the influence of high derivatives terms expected in modifications of the Einstein action. It is well-known that gravity model with terms quadratic in curvature contains pathologies unless fine-tuned \cite{Stelle:1977ry,Accioly:2000nm,Zwiebach:1985uq,Alvarez-Gaume:2015rwa}. Consequently, a great deal of consideration shall be paid before any implementations of higher derivatives/higher curvature gravity are made. Lastly, opportunities to improve performance of the package will be explored. Namely, xAct bundle \cite{xActBundle,Portugal:1998qi,MARTINGARCIA2008597,Brizuela:2008ra,Pitrou:2013hga,Nutma:2013zea} provides a framework to perform symbolic calculations within various gravity models. It includes a tool to obtain explicit perturbative expressions which may significantly improve performance of FeynGrav.

\section*{Acknowledgment}
The work was supported by the Foundation for the Advancement of Theoretical Physics and Mathematics “BASIS”. The author is grateful to A. Arbuzov, A. Bednyakov, D. Kazakov, A. Pikelner, and A. Pimikov for fruitful discussions.

\bibliographystyle{unsrturl}
\bibliography{SFrfGR.bib}

\appendix
\newpage
\section{The complete list of FeynGrav commands}\label{appendix_commands}

let us discuss the complete list of commands defined in FeynGrav. Within the package ``FeynGravCommands'' provides a list of all available commands. We start with commands devoted to interaction vertices which are given in Table \ref{Vertices_Table}.

\begin{table}[ht]
  \begin{center}
    \begin{tabular}{c|c}
      Command & Description\\ \hline
      & \\
      ``GravitonVertex'' & \begin{minipage}{.65\textwidth}The command generates an expression for a graviton interaction vertex with $n\geq 3$ graviton lines. It takes $3n$ arguments $\mu_1,\nu_1,p_1,\cdots,\mu_n,\nu_n,p_n$ where $\mu_i$, $\nu_i$ are graviton lines Lorentz indices, $p_i$ are the corresponding momenta. \end{minipage}\\ &\\ \hline \\ & \\
      ``GravitonScalarVertex'' & \begin{minipage}{.65\textwidth}The command generates an expression for an interaction between $n\geq 1$ gravitons and a scalar field kinetic energy. It takes $2 n + 2$ arguments. The first $2n$ arguments are Lorentz indices of graviton lines. The last two arguments are scalar field momenta.\end{minipage} \\ &\\ \hline & \\
      ``GravitonMassiveScalarVertex'' & \begin{minipage}{.65\textwidth}The command generates an expression for an interaction between $n\geq 1$ gravitons and a kinetic energy of a scalar field with a non-vanishing mass. It takes $2 n + 2 + 1$ arguments. The first $2n$ arguments are Lorentz indices of graviton lines. The other two arguments are scalar field momenta. The last argument is the scalar field mass.\end{minipage} \\ &\\ \hline & \\
      ``GravitonVectorVertex'' & \begin{minipage}{.65\textwidth}The command generates an expression for an interaction between $n\geq 1$ gravitons and a vector field kinetic energy. It takes $2 n + 4$ arguments. The first $2n$ arguments are Lorentz indices of graviton lines. The next two arguments are vector lines Lorentz indices. The last two arguments are corresponding vector field momenta.\end{minipage} \\ &\\ \hline & \\
      ``GravitonFermionVertex'' & \begin{minipage}{.65\textwidth}The command generates an expression for an interaction between $n\geq 1$ gravitons and a Dirac fermion kinetic energy. It takes $2 n + 2$ arguments. The first $2n$ arguments are Lorentz indices of graviton lines. The last two arguments are fermion momenta.\end{minipage} \\ &\\ \hline & \\
      ``GravitonMassiveFermionVertex'' & \begin{minipage}{.65\textwidth}The command generates an expression for an interaction between $n\geq 1$ gravitons and a kinetic energy of a Dirac fermion with a non-vanishing mass. It takes $2 n + 2 + 1$ arguments. The first $2n$ arguments are Lorentz indices of graviton lines. The other two arguments are fermion momenta. The last argument is the Dirac field mass.\end{minipage} \\ &\\ \hline
    \end{tabular}
  \end{center}
  \caption{Description of functions generating interaction vertices.}
  \label{Vertices_Table}
\end{table}

\newpage

Table \ref{Nieuwenhuizen_Table} contains commands related with the standard gauge projectors and Nieuwenhuizen operators \cite{VanNieuwenhuizen:1973fi,Accioly:2000nm,Latosh:2020ysu}. We use the following notation for gauge projectors:
\begin{align}
  \theta_{\mu\nu}(p) &= \eta_{\mu\nu} - \cfrac{p_\mu p_\nu}{p^2}\,,& \omega_{\mu\nu}(p) & = \cfrac{p_\mu p_\nu}{p^2} \, .
\end{align}
In Table \ref{Nieuwenhuizen_Table} we present a correspondence between FeynGrav commands and the discussed operators.
\begin{table}[ht]
  \begin{center}
    \begin{tabular}{c|c}
      Command & Description\\ \hline
      & \\
      ``GaugeProjector'' & \begin{minipage}{.70\textwidth}  $\text{GaugeProjector}[\mu,\nu,p] = \theta_{\mu\nu}(p)$ \end{minipage} \\ & \\ \hline & \\
      ``NieuwenhuizenOperator1'' & \begin{minipage}{.70\textwidth} $\text{NieuwenhuizenOperator1}[\mu,\nu,\alpha,\beta,p]\\ = P^1_{\mu\nu\alpha\beta} = \cfrac12\Big(\theta_{\mu\alpha}(p) \omega_{\nu\beta}(p) + \theta_{\mu\beta}(p) \omega_{\nu\alpha}(p) + \theta_{\nu\alpha}(p) \omega_{\mu\beta}(p) + \theta_{\nu\beta}(p) \omega_{\mu\alpha}(p)\Big)$ \end{minipage} \\ & \\ \hline &\\
      ``NieuwenhuizenOperator2'' & \begin{minipage}{.70\textwidth} $\text{NieuwenhuizenOperator2}[\mu,\nu,\alpha,\beta,p]\\ = P^2_{\mu\nu\alpha\beta}=\cfrac12\Big(\theta_{\mu\alpha}(p)\theta_{\nu\beta}(p) + \theta_{\mu\beta}(p)\theta_{\nu\alpha}(p)\Big)-\cfrac13\Big(\theta_{\mu\nu}(p)\theta_{\alpha\beta}(p)\Big)$ \end{minipage} \\ & \\ \hline &\\
      ``NieuwenhuizenOperator0'' & \begin{minipage}{.70\textwidth} $\text{NieuwenhuizenOperator0}[\mu,\nu,\alpha,\beta,p]\\=P^0_{\mu\nu\alpha\beta}=\cfrac13\Big(\theta_{\mu\nu}(p)\theta_{\alpha\beta}(p)\Big)$ \end{minipage} \\ & \\ \hline &\\
      ``NieuwenhuizenOperator0Bar'' & \begin{minipage}{.70\textwidth} $\text{NieuwenhuizenOperator0Bar}[\mu,\nu,\alpha,\beta,p]\\=\overline{P}_0^{\mu\nu\alpha\beta}=\omega_{\mu\nu}(p)\omega_{\alpha\beta}(p)$ \end{minipage} \\ & \\ \hline &\\
      ``NieuwenhuizenOperator0BarBar'' & \begin{minipage}{.70\textwidth} $\text{NieuwenhuizenOperator0BarBar}[\mu,\nu,\alpha,\beta,p] \\=\overline{\overline{P}}_0^{\mu\nu\alpha\beta}= \theta_{\mu\nu}(p)\omega_{\alpha\beta}(p) + \omega_{\mu\nu}(p)\theta_{\alpha\beta}(p)$ \end{minipage} \\ & \\ \hline
    \end{tabular}
  \end{center}
  \caption{Description of functions generating gauge projectors and Nieuwenhuizen operators.}
  \label{Nieuwenhuizen_Table}
\end{table}

\newpage

\noindent Commands that operate with graviton propagators are described in Table \ref{Propagators_Table}.

\begin{table}[ht]
  \begin{center}
    \begin{tabular}{c|c}
      Command & Description\\ \hline
      & \\
      ``GravitonPropagator'' & \begin{minipage}{.65\textwidth} The command takes five arguments. First four arguments are the graviton line Lorentz indices. The last one is the graviton momentum. The command uses ``FAD'' to deal with the denominator. It is design to better operate with ``TID'' and loop amplitudes. \end{minipage} \\ & \\ \hline &\\
      ``GravitonPropagatorAlternative'' & \begin{minipage}{.65\textwidth} The command takes five arguments. First four arguments are the graviton line Lorentz indices. The last one is the graviton momentum. The command uses ``SPD'' to deal with the denominator. It is design to better operate with tree-level expressions. \end{minipage} \\ & \\ \hline &\\
      ``GravitonPropagatorTop'' & \begin{minipage}{.65\textwidth} The command takes four arguments and returns the following factor that appears at the to of a graviton propagator. $\text{GravitonPropagatorTop}[\mu,\nu,\alpha,\beta] = \cfrac12\left(\eta_{\mu\alpha}\eta_{\nu\beta}+\eta_{\mu\beta}\eta_{\nu\alpha} - \eta_{\mu\nu}\eta_{\alpha\beta}\right)$\end{minipage} \\ & \\ \hline
    \end{tabular}
  \end{center}
  \caption{Description of functions related with the graviton propagator.}
  \label{Propagators_Table}
\end{table}

\newpage

\noindent Lastly, we present a few examples that can help clarify usage of FeynGrav rules in Table \ref{Examples_Table}. We present a few interaction vertices and the corresponding commands. On all diagrams all momenta are directed inwards.

\begin{table}[ht]
  \begin{center}
    \begin{tabular}{c|c}
      Diagram & Command \\ \hline & \\
      \begin{minipage}{.35\textwidth}
        \begin{align*}
          \begin{gathered}
            \begin{fmffile}{D01}
              \begin{fmfgraph*}(30,30)
                \fmfleft{L}
                \fmfright{R1,R2}
                \fmf{dbl_wiggly}{L,V}
                \fmf{dbl_wiggly}{R1,V}
                \fmf{dbl_wiggly}{R2,V}
                \fmfdot{V}
                \fmflabel{$\mu_1\nu_1$,$p_1$}{L}
                \fmflabel{$\mu_2\nu_2$,$p_2$}{R1}
                \fmflabel{$\mu_3\nu_3$,$p_3$}{R2}
              \end{fmfgraph*}
            \end{fmffile}
          \end{gathered}
        \end{align*}
      \end{minipage}
      &
      \begin{minipage}{.55\textwidth}
        \begin{align*}
          \text{GravitonVertex}[\mu_1,\nu_1,p_1,\mu_2,\nu_2,p_2,\mu_3,\nu_3,p_3]
        \end{align*}
      \end{minipage}\\ & \\ \hline & \\
      \begin{minipage}{.35\textwidth}
        \begin{align*}
          \begin{gathered}
            \begin{fmffile}{D02}
              \begin{fmfgraph*}(30,30)
                \fmfleft{L1,L2}
                \fmfright{R1,R2}
                \fmf{dbl_wiggly}{L1,V,L2}
                \fmf{dbl_wiggly}{R1,V,R2}
                \fmfdot{V}
                \fmflabel{$\mu_1\nu_1,p_1$}{L1}
                \fmflabel{$\mu_2\nu_2,p_2$}{L2}
                \fmflabel{$\mu_3\nu_3,p_3$}{R1}
                \fmflabel{$\mu_4\nu_4,p_4$}{R2}
              \end{fmfgraph*}
            \end{fmffile}                
          \end{gathered}
        \end{align*}
      \end{minipage}
      &
      \begin{minipage}{.55\textwidth}
        \begin{align*}
          \text{GravitonVertex}[\mu_1,\nu_1,p_1,\mu_2,\nu_2,p_2,\mu_3,\nu_3,p_3,\mu_4,\nu_4,p_4]
        \end{align*}
      \end{minipage} \\ & \\ \hline & \\
      \begin{minipage}{.35\textwidth}
        \begin{align*}
          \begin{gathered}
            \begin{fmffile}{D03}
              \begin{fmfgraph*}(30,30)
                \fmfleft{L}
                \fmfright{R1,R2}
                \fmf{dbl_wiggly}{L,V}
                \fmf{dashes}{R1,V}
                \fmf{dashes}{R2,V}
                \fmfdot{V}
                \fmflabel{$\mu\nu$}{L}
                \fmflabel{$p_1$}{R1}
                \fmflabel{$p_2$}{R2}
              \end{fmfgraph*}
            \end{fmffile}
          \end{gathered}
        \end{align*}
      \end{minipage}
      &
      \begin{minipage}{.55\textwidth}
        \begin{align*}
          \text{GravitonScalarVertex}[\mu,\nu,p_1,p_2]
        \end{align*}
      \end{minipage}\\ & \\ \hline & \\
      \begin{minipage}{.35\textwidth}
        \begin{align*}
          \begin{gathered}
            \begin{fmffile}{D04}
              \begin{fmfgraph*}(30,30)
                \fmfleft{L1,L2}
                \fmfright{R1,R2}
                \fmf{dbl_wiggly}{L1,V,L2}
                \fmf{dashes}{R1,V}
                \fmf{dashes}{R2,V}
                \fmfdot{V}
                \fmflabel{$\mu_1\nu_1$}{L1}
                \fmflabel{$\mu_2\nu_2$}{L2}
                \fmflabel{$p_1$}{R1}
                \fmflabel{$p_2$}{R2}
              \end{fmfgraph*}
            \end{fmffile}
          \end{gathered}
        \end{align*}
      \end{minipage}
      &
      \begin{minipage}{.55\textwidth}
        \begin{align*}
          \text{GravitonScalarVertex}[\mu_1,\nu_1,\mu_2,\nu_2,p_1,p_2]
        \end{align*}
      \end{minipage} \\ & \\ \hline & \\
      \begin{minipage}{.35\textwidth}
        \begin{align*}
          \begin{gathered}
            \begin{fmffile}{D05}
              \begin{fmfgraph*}(30,30)
                \fmfleft{L}
                \fmfright{R1,R2}
                \fmf{dbl_wiggly}{L,V}
                \fmf{photon}{R1,V}
                \fmf{photon}{R2,V}
                \fmfdot{V}
                \fmflabel{$\mu\nu$}{L}
                \fmflabel{$\alpha,p_1$}{R1}
                \fmflabel{$\beta,p_2$}{R2}
              \end{fmfgraph*}
            \end{fmffile}
          \end{gathered}
        \end{align*}
      \end{minipage}
      &
      \begin{minipage}{.55\textwidth}
        \begin{align*}
          \text{GravitonVectorVertex}[\mu,\nu,\alpha,\beta,p_1,p_2]
        \end{align*}
      \end{minipage} \\ & \\ \hline & \\
      \begin{minipage}{.35\textwidth}
        \begin{align*}
          \begin{gathered}
            \begin{fmffile}{D06}
              \begin{fmfgraph*}(40,40)
                \fmfleft{L}
                \fmfright{R1,R2}
                \fmf{dbl_wiggly}{L,V}
                \fmf{fermion}{R1,V}
                \fmf{fermion}{V,R2}
                \fmfdot{V}
                \fmflabel{$\mu\nu$}{L}
                \fmflabel{$p_1$}{R1}
                \fmflabel{$p_2$}{R2}
              \end{fmfgraph*}
            \end{fmffile}
          \end{gathered}
        \end{align*}
      \end{minipage}
      &
      \begin{minipage}{.55\textwidth}
        \begin{align*}
          \text{GravitonFermionVertex}[\mu,\nu,p_1,p_2]
        \end{align*}
      \end{minipage} \\ & \\ \hline
    \end{tabular}
  \end{center}
  \caption{Examples of FeynGrav commands with the corresponding diagrams. On all diagrams all momenta are directed inwards.}
  \label{Examples_Table}
\end{table}

\newpage
\section{Structure functions}\label{Appendix_Structure_Functions}

\begin{small}
  \begin{align}
    (F_{1,1})_{p_1}^{B_0} =& \cfrac{\kappa^3}{4096}\,\cfrac{1}{(d-2)(d-1)}\,\cfrac{1}{\left[(p_1\cdot p_2)^2 - p_1^2\, p_2^2\right]^2}\,\times
  \end{align}
  \begin{align*}
    \Bigg\{& -2 (d-2) (d^3-8 d^2+24 d-16) k^{10} \\
    & +(d-2) k^8 \Big[ 2 (15 d^3-118 d^2+252 d-144) p_2^2+(d^5-19 d^4+132 d^3-408 d^2+560 d-224) p_1^2 \Big] \\
    & -2 k^6 \Big[2 (d-2) \left(31 d^3-246 d^2+524 d-304\right) p_2^4+2 (d-2) \left(d^5-19 d^4+127 d^3-368 d^2+440 d-144\right) p_1^4\\
      &\hspace{20pt}+\left(d^6-26 d^5+271 d^4-1350 d^3+3308 d^2-3656 d+1376\right) p_2^2 p_1^2\Big]\\
    &+ 2 k^4 \Big[ 2 (d-2) \left(53 d^3-424 d^2+936 d-560\right) p_2^6+(d-2) \left(3 d^5-57 d^4+376 d^3-1064 d^2+1200 d-352\right) p_1^6\\
      &\hspace{20pt}+(d-2) \left(2 d^5-43 d^4+399 d^3-1692 d^2+2988 d-1552\right) p_2^4 p_1^2\\
      &\hspace{20pt}-\left(d^6-6 d^5-51 d^4+562 d^3-1900 d^2+2456 d-992\right) p_2^2 p_1^4 \Big] \\
    & -2 k^2 \Big[(d-2) \left(81 d^3-652 d^2+1488 d-912\right) p_2^8+(d-2) \left(2 d^5-38 d^4+249 d^3-696 d^2+760 d-208\right) p_1^8\\
      &\hspace{20pt}-2 (d-4) \left(2 d^5-27 d^4+152 d^3-392 d^2+412 d-136\right) p_2^4 p_1^4\\
      &\hspace{20pt}+(d-4) \left(3 d^5-46 d^4+281 d^3-782 d^2+988 d-424\right) p_2^6 p_1^2 \\
      &\hspace{20pt}-\left(5 d^6-90 d^5+651 d^4-2334 d^3+4172 d^2-3464 d+992\right) p_2^2 p_1^6\Big]\\
    & +(p_1^2-p_2^2) \Big[ -2 (d-2) \left(23 d^3-186 d^2+436 d-272\right) p_2^8+(d-2) \left(d^5-19 d^4+124 d^3-344 d^2+368 d-96\right) p_1^8\\
      &\hspace{20pt}-(d-2) \left(5 d^5-85 d^4+560 d^3-1676 d^2+2096 d-800\right) p_2^2 p_1^6\\
      &\hspace{20pt}-(d-2) \left(3 d^5-47 d^4+284 d^3-756 d^2+832 d-288\right) p_2^6 p_1^2\\
      &\hspace{20pt}-\left(d^6+7 d^5-136 d^4+592 d^3-1248 d^2+1520 d-640\right) p_2^4 p_1^4 \Big] \Bigg\} .
  \end{align*}
\end{small}

\begin{small}
  \begin{align}
    (F_{1,2})_{p_1}^{B_0} =& \cfrac{\kappa^3}{2048}\,\cfrac{1}{d-2}\,\cfrac{p_1^2}{\left[(p_1\cdot p_2)^2 - p_1^2\, p_2^2\right]^2}\,\times
  \end{align}
  \begin{align*}
    \times\Bigg\{& (d-2) (d^2-8 d+8) (d^2-8 d+24) \, k^8 \\
    &-4 k^6 \Big[(d-2) \left(d^4-16 d^3+100 d^2-284 d+256\right) p_2^2+\left(d^5-18 d^4+128 d^3-452 d^2+760 d-464\right) p_1^2 \Big] \\
    & + 2 k^4 \Big[3 (d-2)^2 \left(d^3-14 d^2+76 d-160\right) p_2^4+(d-4) \left(3 d^4-42 d^3+204 d^2-444 d+328\right) p_1^4\\
      &\hspace{20pt}+2 (d-2) \left(2 d^4-29 d^3+174 d^2-506 d+552\right) p_2^2 p_1^2 \Big]\\
    & - 4 k^2 \Big[\left(2 d^4-25 d^3+116 d^2-204 d+80\right) p_2^2 p_1^4+(d-2) \left(d^4-16 d^3+108 d^2-340 d+384\right) p_2^6\\
      & \hspace{20pt}+\left(d^5-18 d^4+116 d^3-352 d^2+528 d-304\right) p_1^6+\left(d^5-12 d^4+68 d^3-236 d^2+456 d-368\right) p_2^4 p_1^2 \Big]\\
    & + (d-4) (p_1^2-p_2^2) \Big[-(d-2) \left(d^3-8 d+24\right) p_2^4 p_1^2+(d-2)^2 \left(d^2-10 d+4\right) p_1^6\\
      &\hspace{20pt}-(d-2) \left(d^3-12 d^2+64 d-112\right) p_2^6-\left(3 d^4-38 d^3+180 d^2-336 d+144\right) p_2^2 p_1^4 \Big] \Bigg\}.
  \end{align*}
\end{small}

\newpage
\begin{small}
  \begin{align}
    \Big(F_{2,1}\Big)_{p_1}^{B_0} =& \cfrac{\kappa^3}{4096}\,\cfrac{1}{(d-2)(d-1)}\,\cfrac{1}{\left[(p_1\cdot p_2)^2 - p_1^2\, p_2^2\right]^2}\,\times
  \end{align}
  \begin{align*}
    \Bigg\{ &-8 (d-6) (d-2) (d-1)^2 \, k^{10} \\
    & + k^8 \Big[8 (d-2) (d-1) \left(7 d^2-49 d+58\right) p_2^2+\left(d^6-22 d^5+225 d^4-1184 d^3+3048 d^2-3560 d+1504\right) p_1^2 \Big] \\
    & -2 k^6 \Big[ 8 (d-2) (d-1) \left(9 d^2-63 d+86\right) p_2^4+2 (d-4) \left(d^5-18 d^4+127 d^3-416 d^2+578 d-276\right) p_1^4\\
      & \hspace{20pt}+\left(3 d^6-60 d^5+559 d^4-2758 d^3+6936 d^2-8208 d+3552\right) p_2^2 p_1^2\Big] \\
    & + 2 k^4 \Big[ 8 (d-2) (d-1) \left(11 d^2-77 d+114\right) p_2^6+(d-2) \left(3 d^5-56 d^4+439 d^3-1682 d^2+2952 d-1680\right) p_2^2 p_1^4\\
      & \hspace{20pt}+\left(3 d^6-66 d^5+559 d^4-2392 d^3+5468 d^2-6128 d+2592\right) p_1^6\\
      & \hspace{20pt}+2 \left(3 d^6-57 d^5+485 d^4-2197 d^3+5244 d^2-6084 d+2624\right) p_2^4 p_1^2 \Big]\\
    & -2 k^2 \Big[4 (d-2) (d-1) \left(13 d^2-91 d+142\right) p_2^8+4 \left(3 d^5-37 d^4+171 d^3-373 d^2+382 d-152\right) p_2^4 p_1^4\\
      & \hspace{20pt}+2 \left(d^6-22 d^5+177 d^4-704 d^3+1500 d^2-1604 d+664\right) p_1^8\\
      & \hspace{20pt}-\left(3 d^6-56 d^5+431 d^4-1682 d^3+3328 d^2-3056 d+1056\right) p_2^2 p_1^6\\
      & \hspace{20pt}+\left(5 d^6-92 d^5+713 d^4-2882 d^3+6264 d^2-6832 d+2848\right) p_2^6 p_1^2\Big]\\
    &+(p_1^2-p_2^2)\Big[-8 (d-2) (d-1) \left(3 d^2-21 d+34\right) p_2^8+(d-4) (d-2) \left(d^4-16 d^3+65 d^2-106 d+60\right) p_1^8\\
      &\hspace{20pt}-(d-2) \left(3 d^5-48 d^4+307 d^3-934 d^2+1332 d-672\right) p_2^6 p_1^2\\
      &\hspace{20pt}-\left(5 d^6-94 d^5+733 d^4-2904 d^3+5952 d^2-5832 d+2176\right) p_2^2 p_1^6\\
      &\hspace{20pt}-\left(d^6+6 d^5-135 d^4+668 d^3-1656 d^2+2232 d-1152\right) p_2^4 p_1^4\Big] \Bigg\}.
  \end{align*}
\end{small}

\begin{small}
  \begin{align}
    \big(F_3\big)_{p_1}^{B_0} =&\cfrac{\kappa^3}{2048}\,\cfrac{1}{(d-2)(d-1)}\,\cfrac{1}{(p_1\cdot p_2)^2 - p_1^2\, p_2^2}\,\times
  \end{align}
  \begin{align*}
    \Bigg\{ & 8 (d-6) (d-2) (d-1)^2 \,k^8 \\
    &+ k^6 \Big[ -16 (d-2) (d-1) \left(3 d^2-21 d+26\right) p_2^2-\left(3 d^6-58 d^5+467 d^4-1968 d^3+4360 d^2-4664 d+1888\right) p_1^2 \Big]\\
    & + k^4 \Big[ (d-2) \left(7 d^5-120 d^4+847 d^3-2958 d^2+4780 d-2640\right) p_2^2 p_1^2\\
      &\hspace{20pt}+\left(7 d^6-136 d^5+1061 d^4-4296 d^3+9424 d^2-10360 d+4448\right) p_1^4+96 (d-5) (d-2)^2 (d-1) p_2^4 \Big] \\
    & + k^2 \Big[ -16 (d-2) (d-1) \left(5 d^2-35 d+54\right) p_2^6-\left(5 d^6-98 d^5+753 d^4-3008 d^3+6696 d^2-7752 d+3616\right) p_1^6\\
      & \hspace{20pt}+2 \left(d^6-18 d^5+119 d^4-330 d^3+216 d^2+528 d-608\right) p_2^2 p_1^4\\
      & \hspace{20pt}-\left(5 d^6-94 d^5+709  d^4-2744 d^3+5832 d^2-6472 d+2848\right) p_2^4 p_1^2 \Big]\\
    & + (p_1^2-p_2^2) \Big[-8 (d-2) (d-1) \left(3 d^2-21 d+34\right) p_2^6-(d-2) d \left(d^4-16 d^3+81 d^2-138 d+44\right) p_2^4 p_1^2\\
      & \hspace{20pt}+(d-2) \left(d^5-18 d^4+115 d^3-370 d^2+660 d-480\right) p_1^6\\
      & \hspace{20pt}-2 \left(d^5-21 d^4+156 d^3-536 d^2+832 d-400\right) p_2^2 p_1^4 \Big]\Bigg\}.
  \end{align*}
\end{small}

\newpage
\begin{small}
  \begin{align}
    \big( F_{1,1}\big)_{k}^{B_0}=&\cfrac{\kappa^3}{8192}\,\cfrac{1}{(d-2)(d-1)}\,\cfrac{1}{\left[(p_1\cdot p_2)^2 - p_1^2\, p_2^2\right]^2}\, \times
  \end{align}
  \begin{align*}
    \Bigg\{ & -(d-2) \left(d^5-23 d^4+214 d^3-748 d^2+888 d-224\right)k^{10} \\
    & + k^8 \Big[(d-2) \left(3 d^5-77 d^4+826 d^3-3084 d^2+3800 d-992\right) p_1^2\\
      & \hspace{20pt}+\left(3 d^6-83 d^5+980 d^4-5024 d^3+11984 d^2-12112 d+3520\right) p_2^2 \Big]\\
    & -2 k^6 \Big[ (d-2) \left(d^5-39 d^4+582 d^3-2428 d^2+3160 d-864\right) p_1^4\\
      & \hspace{20pt}-2 \left(7 d^5-229 d^4+1718 d^3-5116 d^2+5960 d-1888\right) p_2^2 p_1^2\\
      & \hspace{20pt}+\left(d^6-41 d^5+644 d^4-3976 d^3+11008 d^2-12464 d+3904\right) p_2^4 \Big]\\
    & -2 k^4 \Big[(d-2) \left(d^5+d^4-338 d^3+1772 d^2-2520 d+736\right) p_1^6\\
      &\hspace{20pt}+(d-2) \left(9 d^5-141 d^4+796 d^3-1756 d^2+960 d-384\right) p_2^4 p_1^2\\
      &\hspace{20pt}+\left(d^6-39 d^5+222 d^4+436 d^3-4936 d^2+8672 d-3584\right) p_2^2 p_1^4\\
      &\hspace{20pt}+\left(d^6-d^5-340 d^4+3216 d^3-10800 d^2+13712 d-4672\right) p_2^6 \Big]\Bigg\}.
  \end{align*}
\end{small}

\begin{small}
  \begin{align}
    \big( F_{2,1}\big)_{k}^{B_0}=&\cfrac{\kappa^3}{8192}\,\cfrac{1}{(d-2)(d-1)}\,\cfrac{1}{\left[(p_1\cdot p_2)^2 - p_1^2\, p_2^2\right]^2} \,\times
  \end{align}
  \begin{align*}
    \Bigg\{ & (-d^6+25 d^5-260 d^4+1272 d^3-3056 d^2+2960 d-576)\,k^{10} \\
    & + (3 d^6-83 d^5+964 d^4-5072 d^3+12624 d^2-12432 d+2368)\,k^8\,(p_1^2 + p_2^2)\\
    & -2 k^6 \Big[\left(d^6-41 d^5+652 d^4-4024 d^3+10872 d^2-11392 d+2496\right) p_1^4\\
      &\hspace{20pt}-2 \left(2 d^6-23 d^5-31 d^4+1076 d^3-4172 d^2+5200 d-1344\right) p_2^2 p_1^2\\
      &\hspace{20pt}+\left(d^6-41 d^5+652 d^4-4024 d^3+10872 d^2-11392 d+2496\right) p_2^4 \Big]\\
    & -2 k^4 (p_1^2+p_2^2) \Big[ \left(d^6-d^5-364 d^4+3216 d^3-10024 d^2+11680 d-3264\right) (p_1^4+p_2^4)\\
      &\hspace{20pt}+2 \left(4 d^6-79 d^5+669 d^4-2738 d^3+5228 d^2-3736 d+96\right) p_2^2 p_1^2 \Big]\\
    & + k^2 \Big[ (d-2) \left(3 d^5-37 d^4-174 d^3+2300 d^2-5480 d+2336\right) (p_1^8+p_2^8)\\
      &\hspace{20pt}-4 \left(7 d^5-135 d^4+768 d^3-1668 d^2+1008 d+384\right) (p_2^2 p_1^6+p_1^2 p_2^6)\\
      &\hspace{20pt}-2 \left(3 d^6-87 d^5+664 d^4-1912 d^3+1904 d^2+368 d-1344\right) p_2^4 p_1^4 \Big]\\
    & - (d-2) \left(p_1^2-p_2^2\right)^2 \left(p_1^2+p_2^2\right) \Big[\left(d^5-15 d^4-2 d^3+460 d^2-1288 d+672\right) (p_1^4+p_2^4)\\
      & \hspace{20pt}-2 \left(3 d^5-43 d^4+184 d^3-204 d^2-176 d+64\right) p_2^2 p_1^2 \Big] \Bigg\}.
  \end{align*}
\end{small}

\newpage
\begin{small}
  \begin{align}
    \big(F_3\big)_{k}^{B_0} = &\cfrac{\kappa^3}{2048}\,\cfrac{1}{(d-2)(d-1)}\,\cfrac{1}{(p_1\cdot p_2)^2 - p_1^2\, p_2^2}\,\times
  \end{align}
  \begin{align*}
    \Bigg\{ & d \left(d^5-23 d^4+226 d^3-1008 d^2+2072 d-1552\right) \, k^{10}\\
    & -d \left(d^5-31 d^4+398 d^3-2092 d^2+4824 d-3936\right) \,k^6\,\left(p_1^2+p_2^2\right) \\
    & +k^4 \Big[-2 (d-2) \left(3 d^5-45 d^4+220 d^3-354 d^2-12 d-16\right) p_2^2 p_1^2\\
      & \hspace{20pt}-\left(d^6-7 d^5-130 d^4+1356 d^3-4304 d^2+4592 d-640\right) (p_1^4+p_2^4) \Big]\\
    & + k^2 \,(p_1^2+p_2^2) \,\Big[ (d-2) \left(d^5-13 d^4+4 d^3+476 d^2-1472 d+640\right) (p_1^4+p_2^4)\\
      &\hspace{20pt} -2 \left(d^6-17 d^5+74 d^4+106 d^3-1088 d^2+1448 d-160\right) p_2^2 p_1^2 \Big]\\
    & +4 (d-2) \left(p_1^2-p_2^2\right)^2 \Big[\left(3 d^3-43 d^2+132 d-80\right) (p_1^4+p_2^4)-2 \left(d^3-25 d^2+84 d-48\right) p_2^2 p_1^2\Big] \Bigg\}.
  \end{align*}
\end{small}

\begin{small}
  \begin{align}
    \big( F_{1,1}\big)^{C_0}=&\cfrac{\kappa^3}{2048}\,\cfrac{1}{d-2}\,\cfrac{p_2^2}{\left[(p_1\cdot p_2)^2 - p_1^2\, p_2^2\right]^2}\,\times
  \end{align}
  \begin{align*}
    \Bigg\{ & 32 (d-2) k^{10} - (d-2) \,k^8\,\Big[4 \left(d^2-6 d+32\right) p_2^2+\left(d^4-16 d^3+88 d^2-200 d+224\right) p_1^2\Big] \\
    & +2 k^6\Big[2 (d-2) \left(7 d^2-46 d+96\right) p_2^4+2 (d-4) (d-2) \left(d^3-12 d^2+44 d-52\right) p_1^4\\
      &\hspace{20pt}+\left(d^5-21 d^4+156 d^3-540 d^2+944 d-608\right) p_2^2 p_1^2\Big]\\
    & -2 k^4 \Big[ 2 (d-2) \left(15 d^2-102 d+176\right) p_2^6+(d-2) \left(3 d^4-48 d^3+288 d^2-768 d+736\right) p_1^6\\
      &\hspace{20pt}+\left(d^5-21 d^4+164 d^3-636 d^2+1264 d-928\right) p_2^2 p_1^4\\
      &\hspace{20pt}+\left(d^5-24 d^4+192 d^3-728 d^2+1376 d-960\right) p_2^4 p_1^2 \Big]\\
    & +2 k^2 \Big[2 (d-2) \left(d^4-16 d^3+100 d^2-284 d+296\right) p_1^8-(d-2) \left(d^4-19 d^3+118 d^2-280 d+176\right) p_2^2 p_1^6\\
      &\hspace{20pt}+(d-2) \left(d^4-19 d^3+102 d^2-216 d+176\right) p_2^6 p_1^2-2 \left(d^5-16 d^4+111 d^3-392 d^2+628 d-288\right) p_2^4 p_1^4\\
      &\hspace{20pt}+2 (d-4) (d-2) (13 d-38) p_2^8 \Big]\\
    & -(d-2)(d-4)\left(p_1^2-p_2^2\right)^2 \Big[ (d-2) \left(d^2-10 d+20\right) p_2^4 p_1^2+\left(d^3-12 d^2+56 d-88\right) p_1^6\\
      &\hspace{20pt}+6 (d-2)^2 p_2^2 p_1^4+16 (d-3) p_2^6 \Big] \Bigg\}.
  \end{align*}
\end{small}

\begin{small}
  \begin{align}
    \big( F_{2,1}\big)^{C_0}=&\cfrac{\kappa^3}{2048}\,\cfrac{1}{d-2}\,\cfrac{1}{\left[(p_1\cdot p_2)^2 - p_1^2\, p_2^2\right]^2}\,\times
  \end{align}
  \begin{align*}
    \Bigg\{ & -16\,(d-2)\,k^{12}+64\,(d-2)\,k^{10}\,(p_1^2+p-2^2) \\
    & + k^8 \Big[-8 (d-2) \left(d^2-7 d+24\right) (p_1^4+p_2^4)+\left(3 d^4-34 d^3+156 d^2-504 d+544\right) p_2^2 p_1^2\Big] \\
    & +k^6\,(p_1^2+p_2^2)\,\Big[ 32 (d-2) \left(d^2-7 d+14\right) (p_1^4+p_2^4) +\left(d^5-27 d^4+214 d^3-780 d^2+1464 d-992\right) p_2^2 p_1^2 \Big]\\
    & +k^4\,\Big[-16 (d-2) \left(3 d^2-21 d+37\right) (p_1^8+p_2^8) +4 (d-2) \left(4 d^3-41 d^2+158 d-256\right) p_2^4 p_1^4 \\
      &\hspace{20pt}-(d-4) \left(3 d^4-51 d^3+298 d^2-780 d+696\right) (p_1^2 p_2^6+p_2^2 p_1^6)\Big] \\
    & + k^2 \,(p_1^2+p_2^2) \Big[ 32 (d-4) (d-3) (d-2) (p_1^8+p_2^8) + (d-2) \left(3 d^4-51 d^3+268 d^2-540 d+336\right) (p_1^2 p_2^6+p_2^2 p_1^6) \\
      &\hspace{20pt} -2 \left(3 d^5-55 d^4+376 d^3-1204 d^2+1744 d-832\right) p_2^4 p_1^4  \Big]\\
    & -(d-4) (d-2) \left(p_1^2-p_2^2\right)^2 \Big[ 8 (d-3) (p_1^8+p_2^8)+(d-4)^3 (p_1^2 p_2^6+p_2^2 p_1^6)+6 (d-2)^2 p_2^4 p_1^4 \Big] \Bigg\}.
  \end{align*}
\end{small}

\begin{small}
  \begin{align}
    \big(F_3\big)^{C_0} = &\cfrac{\kappa^3}{1024}\,\cfrac{1}{d-2}\,\cfrac{1}{(p_1\cdot p_2)^2 - p_1^2\, p_2^2}\,\times
  \end{align}
  \begin{align*}
    \Bigg\{ & 16 (d-2)k^{10} - 48 (d-2) \,k^8\, (p_1^2+p_2^2) \\
    & + k^6 \Big[ 8 (d-2) \left(d^2-7 d+18\right) (p_1^4+p_2^4)+\left(d^5-19 d^4+134 d^3-460 d^2+824 d-544\right) p_2^2 p_1^2 \Big]\\
    & -2 \,k^4\,(p_1^2+p_2^2)\,\Big[ 4 (d-2) \left(3 d^2-21 d+38\right) (p_1^4+p_2^4) +(d-4) \left(d^4-15 d^3+68 d^2-140 d+96\right) p_2^2 p_1^2 \Big] \\
    & +(d-4)\,k^2\,\Big[ 24 (d-3) (d-2) (p_1^8+p_2^8) +(d-2) \left(d^3-13 d^2+40 d-60\right) (p_1^2 p_2^6 + p_2^2 p_1^6)\\
      &\hspace{20pt} -2 (d-4) \left(d^3-9 d^2+20 d+4\right) p_2^4 p_1^4 \Big] \\
    & -4 (d-4) (d-2) \left(p_1^2-p_2^2\right)^2 \left(p_1^2+p_2^2\right) \Big[ 2 (d-3) (p_1^4+p_2^4) -(3 d-8) p_2^2 p_1^2\Big]
    \Bigg\}.
  \end{align*}
\end{small}

\end{document}